\documentclass{aa}
\usepackage{graphicx}
\usepackage{txfonts}
\usepackage{xcolor}

\usepackage{placeins}

\usepackage[colorlinks=true,
            linkcolor=red,
            urlcolor=blue,
            citecolor=blue]{hyperref}

\usepackage[normalem]{ulem}

\begin{document}

   \title{The eROSITA Final Equatorial-Depth Survey (eFEDS):}
   
    \subtitle{The variability catalogue and multi-epoch comparison}

\titlerunning{eFEDS variability}

   \authorrunning{Th. Boller, et al.}

   \author{Th. Boller\inst{1}\thanks{E-mail: bol@mpe.mpg.de}, J.H.M.M. Schmitt\inst{2}, J. Buchner\inst{1}, M. Freyberg\inst{1}, A. Georgakakis\inst{3}, T. Liu\inst{1}, J. Robrade\inst{2}, A. Merloni\inst{1}, K.~Nandra\inst{1}, A. Malyali\inst{1}, M. Krumpe\inst{4}, M. Salvato\inst{1}, and T. Dwelly\inst{1}
          }

   \institute{Max-Planck-Institut f\"ur extraterrestrische Physik, Giessenbachstrasse 1, 85748 Garching, Germany
   \and
   Hamburger Sternwarte, Universit\"at Hamburg, Gojenbergsweg 112,  21029 Hamburg, Germany
   \and
   National Observatory of Athens, I. Metaxa \& V. Pavlou, P. Penteli, 15236, Athens, Greece
            \and
   Leibniz-Institut für Astrophysik Potsdam (AIP), An der  Sternwarte 16, 14482 Potsdam, Germany
             }

   \date{Received April 22, 2021; accepted June 21, 2021}

  \abstract{
   {The 140 square degree Final Equatorial-Depth Survey (eFEDS) field, observed with the extended ROentgen Survey with an Imaging Telescope Array (eROSITA) 
   aboard the Spectrum-Roentgen-Gamma (SRG) mission,
   provides a first look at the variable eROSITA sky.  
   We analyze the intrinsic X-ray variability of the eFEDS sources, provide X-ray light curves and tables with variability test results in the 0.2-2.3 keV (soft) and
   2.3-5.0 keV (hard) bands.
   }
   {We perform variability tests using the normalized excess variance and maximum amplitude variability methods as performed for the 2RXS catalogue and add results from the Bayesian excess variance and the Bayesian block methods. In total 65 sources have been identified as being significantly variable in the soft band. In the hard band only one source is found to vary significantly.
}
   {For the most variable sources fits to stellar flare events reveal extreme flare properties. 
   A few highly variable AGN have also been detected. About half of the variable eFEDS sources have been detected at X-rays with eROSITA for the first time. Comparison with 2RXS and XMM observations provide variability information on timescales of years to decades.
   }
   {}
}


\maketitle
%

\section{Introduction}

Variability is a ubiquitous and defining characteristic of coronal stellar and accreting compact object systems, which represent two major object classes of variable X-ray sources.  Typically, the most rapid and largest amplitude variations are observed in the X-ray band, suggesting that the X-rays originate from magnetic reconnection processes in stars or that they arise very close to the compact object, as do the processes responsible for the variability. Observational studies of stellar X-ray activity and of accreting supermassive black hole systems have revealed a rich phenomenology in terms of observations, and associated theoretical models to explain them. 

The eFEDS (eROSITA Final Equatorial-Depth Survey) field was observed in November 2019 by SRG/eROSITA \citep{2021Predehl} as part of its Performance Verification programme and was executed in a so-called field scan mode. Further details and a catalogue of X-ray sources can be found in \cite{Brunner2021}.
The eFEDS field provides a deep view of X-ray variability over an unprecedentedly large, contiguous field of about 140 square degrees on timescales from seconds, to hours, and even longer timescales when comparing to previous observations by other X-ray observatories. For example, comparison of eFEDS sources with XMM-Newton observations provides information on source variability on time scales of years. The 2RXS catalogue \citep{2016Boller} contains detailed variability information from an even earlier epoch, which is useful to compare with the timing properties of sources detected in the eFEDS field. 

\section{Observations and data preparation}

The eFEDS field scan rate is 0.003654 deg/s, roughly 1/7 eROSITA's all-sky survey scan rate. The resulting field of view (FOV) passing time of a given source is about 300 seconds, and each position is covered again typically 
after about 2000 seconds.
Typically 3 consecutive visits are made.
For sources at the eFEDS field borders there are scan reversals and more than 3 consecutive visits are available as such sources cross the FOV twice.  
The total elapsed time from the first to the last observation of a source is up to about 22000 seconds, with the precise temporal sampling depending on the position within the field.

The eFEDS survey scanning strategy is especially sensitive to stellar flaring events, which have decay times of the order of a few thousand seconds. This can be compared to the eROSITA survey mode where only one data point is collected every 14000 seconds for about 7 passes with that interval followed by a large time gap, depending on where on the sky the source is.
The eFEDS data also contrast with the ROSAT all-sky survey where data points were obtained every 5760 seconds, however on a much longer timescale of about two days.

The eFEDS event file is processed and cleaned as described in \citet{Brunner2021}.
We extract the light curves from the cleaned event file for all the 27910 sources in the eFEDS main X-ray catalog \citep{Brunner2021}
using the eROSITA Science Analysis Software System (eSASS) task \texttt{srctool}~\footnote{https://erosita.mpe.mpg.de/eROdoc/tasks/srctool\_doc.html\#Light-curve\_file(s)} (Version 1.63; \cite{Brunner2021}).
The source extraction region is defined as a circular region that maximizes the source signal-to-noise ratio.
The background extraction region is defined as an annular region that is 200 times larger than the source extraction region in area. Contamination from nearby sources are excluded from the source and background regions. More details about the regions are described in \cite{Liu2021},
where the same regions are used to extract the X-ray spectra.
With these regions input to \texttt{srctool}, we extract the light curves in the (0.2-5), (0.2-2.3), and (2.3-5) keV bands, adopting all the valid event patterns ($PATTERN<=15$) and a time bin size of $\Delta t=100$ seconds.
The \texttt{srctool} measures the background counts $B$, source counts $S$, the scaling factor $r$ between the source and background regions, and then calculates the net source count rate $X_i$ in each time bin $i$ as 
$$X_i=\frac{S_i-B_i\times r}{f_{\mathrm{exp},i}\Delta t}$$
The effective exposure fraction $f_\mathrm{exp}$ is calculated to account for the instrumental factors that affect the source photon counts in each time bin and each energy band at a given source count rate. It is the product of the fraction of the time bin which overlaps with input Good time Intervals (GTIs) and the fraction of the nominal effective collecting area seen by the source, which accounts for both the local telescope vignetting and the flux loss caused by existence of bad pixels, exceeding the boundary of the instrumental FOV, and PSF-wing outside the source extraction region.
The \texttt{srctool} uses the total counts in the source and background regions to calculate the count rate errors as the square root of the counts. 
In the cases of low counts ($<25$), this method does not work and thus no error is provided by \texttt{srctool}. In such cases, we estimate the counts error as $1+\sqrt{\mathrm{counts}+0.75}$ \citep{1986Gehrels}. 
This is important as otherwise such time bins would have been excluded from the variability analysis, which would bias the analysis.

\section{Methodology}

\subsection{eROSITA variability tests}

In the following we analyse the variability properties of the eFEDS point sources. In each of the three energy bands, four different variability characterization methods are applied. These include the normalized excess variance (NEV) \citep{Edelson1990,Nandra1997,Edelson2002}, maximum amplitude variability \citep{2016Boller}, Bayesian blocks \citep{Scargle2013} and the Bayesian excess variance 
\citep{Buchner2021}.
The tests are presented in detail in \cite{Buchner2021}, who also study their performance for eFEDS surveys in detail. Here, we give only a brief introduction to each method.

\subsubsection{Maximum amplitude variability}

The maximum amplitude variability is defined as the span between the most extreme values of the count rate:

\begin{equation}\label{eq:amplmax}
ampl\_max = (X_\mathrm{max} - \sigma_\mathrm{max})  -  (X_\mathrm{min} + \sigma_\mathrm{min})
\end{equation}
where $X_\mathrm{max}$ ($X_\mathrm{min}$) is the maximum (minimum) count rate and the associated error is $\sigma_\mathrm{max}$ ($\sigma_\mathrm{min}$).
The uncertainty is:
\begin{equation}\label{eq:sigma_amplmax}
\sigma(ampl\_max) = \sqrt{\sigma_\mathrm{max}^2 + \sigma_\mathrm{min}^2}
\end{equation}

The ratio of $ampl\_max$ to $\sigma(ampl\_max)$ gives the significance of the maximum amplitude variability in units of $\sigma$. \cite{Buchner2021} identified with simulations that at a threshold of $ampl\_max / \sigma(ampl\_max) > 2.6$ we expect no false positives in an eFEDS-like field.
The definition of ampl\_max is a conservative lower limit on the variability and follows its original definition in \cite{2016Boller}.

\subsubsection{Excess variability}

The NEV is defined as the difference between the expected variance from the error bars $\sigma_i$ and the observed variance:
\begin{equation}\label{eq:NEV}
NEV = \frac{<(X_i - \mu)^2> - <\sigma_i^2>}{\mu^2}
\end{equation}
The NEV estimator does not have a analytic uncertainty. \cite{Vaughan2003} performed extensive numerical studies and found the empirical formula:
\begin{equation}\label{eq:sigma_NEV}
\sigma(NEV) =  \sqrt{\frac{2}{N} (\frac{<\sigma_i^2>}{\mu^2})^2 + \frac{<\sigma_i^2>}{N} (\frac{2F_{var}}{\mu})^2}
\end{equation}
where $\mu$ is the mean count rate,
$\sigma_i$ are the count rate uncertainties, and
$<\sigma_i^2>$ is the mean of the square of the count rate uncertainties.
N is the number of data points. The fractional variability $F_{var}$ \citep{Edelson1990} is simply the square root of the NEV defined by \cite{Nandra1997}.
The ratio of $NEV$ and $\sigma(NEV)$ gives the significance of the normalized excess variance in units of $\sigma$. The simulations of \cite{Buchner2021} suggest that a threshold of $NEV / \sigma(NEV) > 1.7$ is appropriate if we wish to avoid false positives in an eFEDS-like field.

The Gaussianity assumption in the calculation of the NEV breaks down in the low count rate regime. \cite{Buchner2021} therefore developed a Bayesian excess variance estimate (bexvar), which works with the Poisson counts instead of inferred uncertainties. For the source variability, the Bayesian excess variance model assumes a log-normal distribution for the counts and infers its variance ($\sigma_\mathrm{bexvar}^2$). 
We use the \texttt{bexvar} Python implementation\footnote{\url{https://github.com/JohannesBuchner/bexvar/}} which computes posterior probability distributions using the nested sampling inference algorithm MLFriends \citep{Buchner2014stats,Buchner2019c} implemented in the \texttt{UltraNest} Python package\footnote{\url{https://johannesbuchner.github.io/UltraNest/}} \citep{Buchner2021b}.
\cite{Buchner2021}
identified that a threshold of $0.14$~dex on the lower 10\% quantile of $\sigma_\mathrm{bexvar}$ has fewer than $0.3\%$ 
false positives
in  eFEDS-like simulated light curves.
The log-normal assumption is a simple distribution, which is guaranteed to always produce positive count rates (unlike, e.g., a normal distribution). Given the few data points per source (20), and low counts in all but a few sources, it is difficult to estimate higher moments than the variance.

\subsubsection{Bayesian blocks}

The Bayesian blocks algorithm \citep{Scargle2013} breaks light curves into segments with constant count rates. 
Starting from the first data point, every iteration of the algorithm adds one data point at a time, making use of stored results of previous iteration to obtain a new exact optimum at each step. The result is the exact global optimum in time of order $N^2$ (which would be impossible with an explicit search of the exponential search space -- $2^N$ in size). In effect it yields the step-function optimally "fitting" the data, from among all possible such representations.
Variability is detected with this method when the light curve was broken into at least two segments.

Bayesian blocks can be applied to photon counts directly, without requiring binning. However, Bayesian blocks applied in such a way could detect variability caused by the varying sensitivity and background. This is because the permanently changing orientation of eROSITA modulates the count rates from the source and some components of the background, which can dominate the signal. To measure the source flux variability (instead of the observed count rate variability), the background and instrument sensitivity at any time needs to be incorporated. While the original count event formulation of Bayesian blocks can incorporate varying sensitivity, contributions from a potentially varying background deduced by an 'off' region 
goes beyond the current work \citep[but see ][for Bayesian blocks extensions]{2019Kerr}
However, when using the Gaussian approximation in a pre-binned light curve, one can infer the (background-corrected) source count rate at each time bin, ($X_i$, $\sigma_i$). The Bayesian block formulation using the Gaussian error bars can then attempt to detect changes. 
When applying these Gaussian Bayesian blocks to eFEDS-like simulations, \citep{Buchner2021} found that this method produces virtually no false positives. We use the Bayesian blocks implementation from astropy \citep{astropy:2013,astropy:2018}\footnote{\url{https://www.astropy.org/}}.
%

\subsection{Variability test comparison}

NEV is testing variability over the length of the observation. It is sensitive to variability trends on that time scale.
Maximum amplitude variability is used to search for flaring events based on the highest and lowest data points.
Bayesian blocks and Bayesian excess variance are in addition sensitive for variability on larger time scales than 100 seconds and in the lower count rate regime.

\subsection{Optical contamination by bright stars}

Bright stars with $g < 4.5$~mag that are detected exclusively in the soft band are likely to be particularly problematic, with the recorded signal being prone to optical contamination. These stars are causing apparent variability and trigger false positives and have been therefore been removed from the sample. 
In a paper appearing in the eROSITA paper splash by
\cite{Schmitt2021} the issue of optical contamination is discussed in detail.

\subsection{Visualisations}

%

Two visualisations are employed: The time series of $X_i$ is plotted with error bars. This allows the identification of strong trends and bright flares. However, because most bins contain few counts, and subsequently the error bars are large, it is difficult to accumulate the evidence of adjacent bins. Therefore, we also use a plot of the cumulative total counts and compare it to the expectation from a constant source with the same net count rate under the same observing conditions.

We classify sources as follows:
\begin{itemize}
\item Non-candidates: if no variability test is triggered. These are the vast majority and not presented in this paper.
\item Potential light leak false positives: if a test triggered, but there is a bright ($g < 4.5$) Gaia source in the close vicinity. 
\item Likely Variables: 
if a test triggered, but the visual inspection of the cumulative counts does not exceed the $2\sigma$ range at any time bin.
\item Secure Variables: if a test triggered, and the visual inspection of the cumulative counts exceeds the $2\sigma$ range at some time bin.
\end{itemize}
The vast majority of eROSITA point sources are either AGN or coronally active stars, and hence are likely to be intrinsically variable. Here we report only the sources where we can detect significant variability based on the eFEDS observations. This will naturally highlight the most extremely variable sources during the observation period.

\section{Results}

\subsection{Variability test results}

\begin{figure}
\includegraphics[width=\columnwidth]{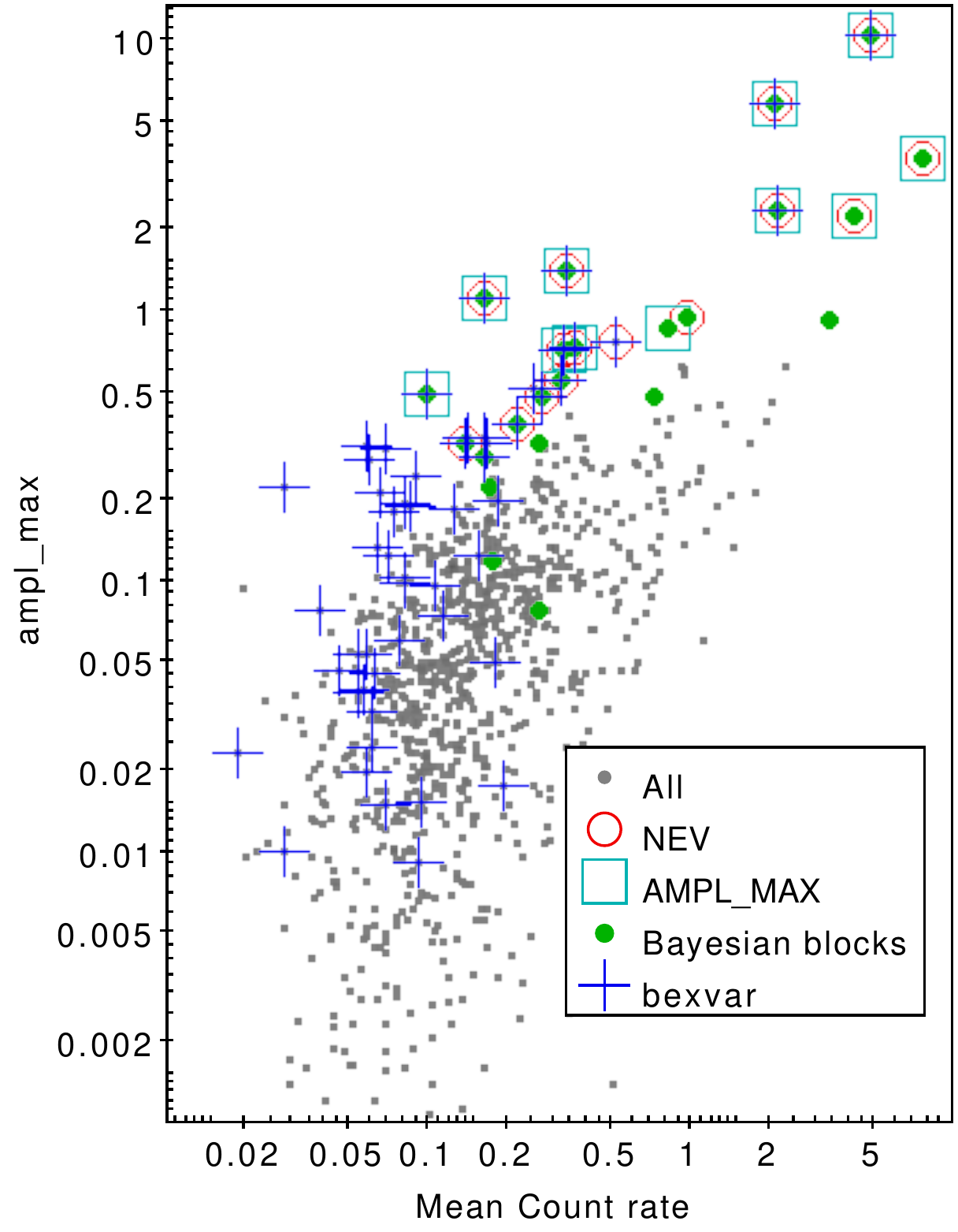}
    \caption{Sample distribution in mean count rate and ampl\_max. For each test, the sources considered significantly variable are marked. While most such markers are limited to the upper left, the blue crosses extend to the bottom left of the plot, reflecting the sensitivity of bexvar in the low count rate regime.}
    \label{fig:sampledist}
\end{figure}

\begin{figure}[!htb]
\includegraphics[height=5.5cm]{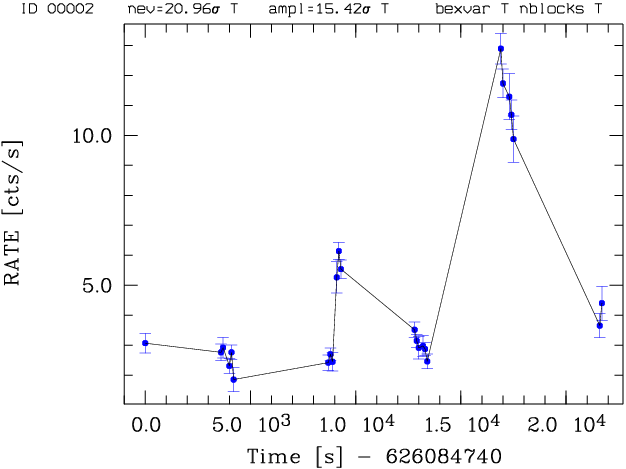}
\includegraphics[height=5.5cm]{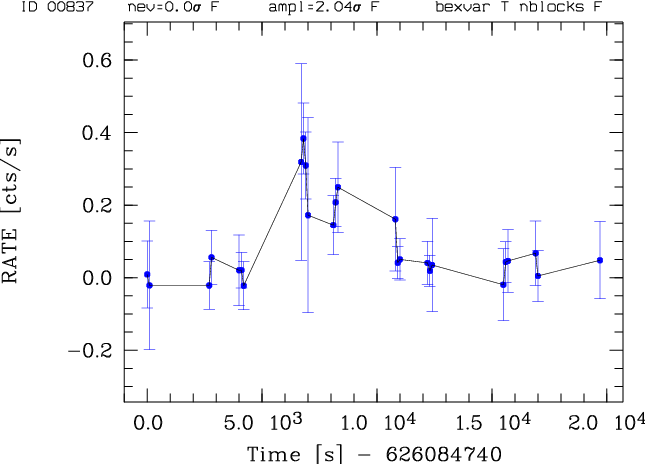}
\includegraphics[height=5.5cm]{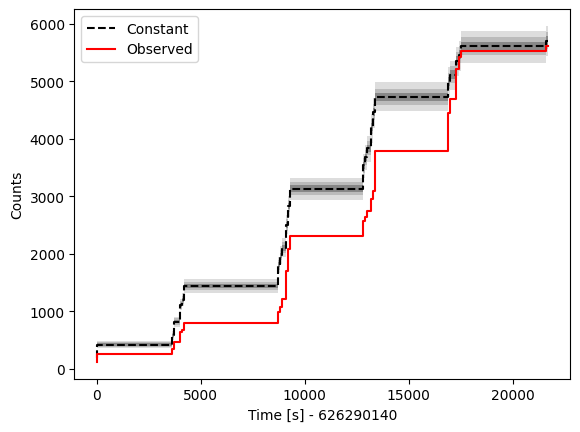}
\includegraphics[height=5.5cm]{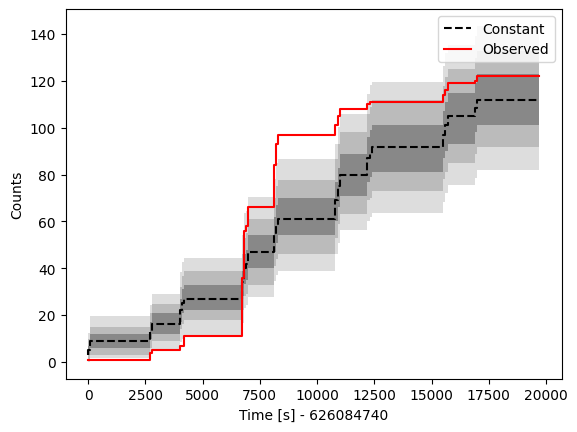}
    \caption{Light curves (top two panels) and cumulative count plots (lower two panels) for source identification 2 and 837, respectively. The grey shadowed areas in the cumulative plots indicate count rate deviations of 1, 2 and 3 $\sigma$ with respect to a constant count rate per cumulative time bin. 
}
    \label{fig:3sigmacounts}
\end{figure}

\begin{figure*}
    \includegraphics[height=4.7cm]{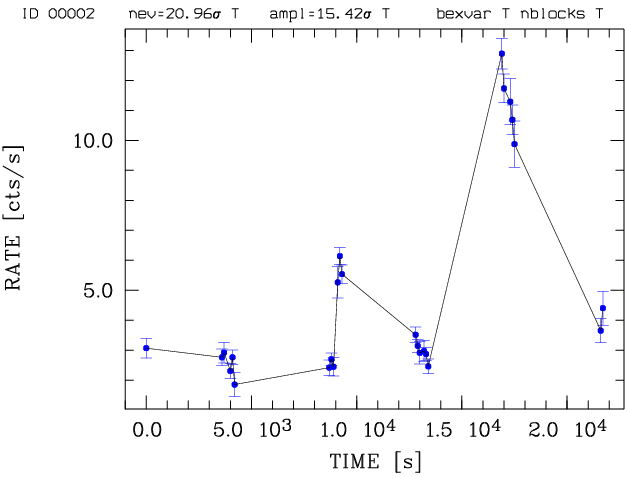}
    \includegraphics[height=4.7cm]{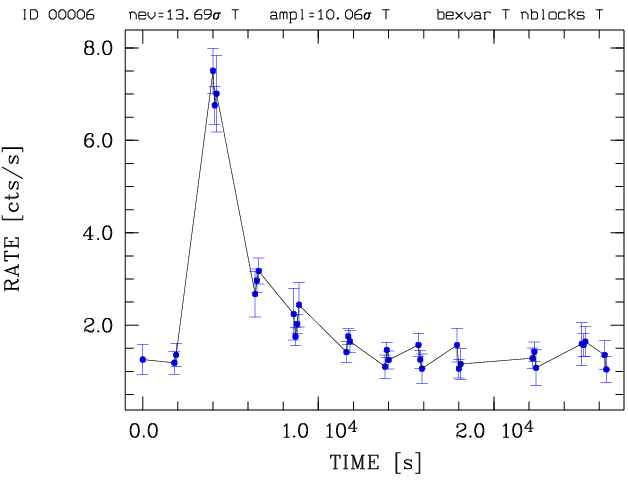}
    \includegraphics[height=4.7cm]{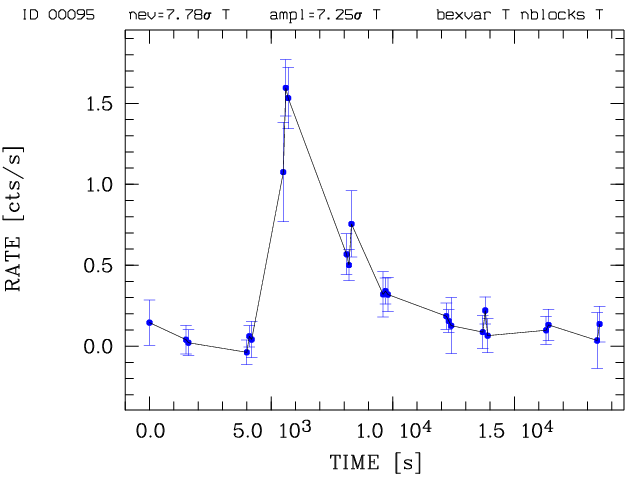}
    \includegraphics[height=4.7cm]{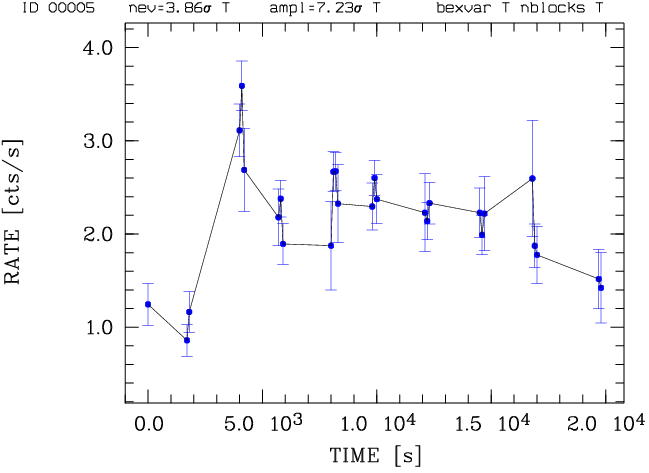}
    \includegraphics[height=4.7cm]{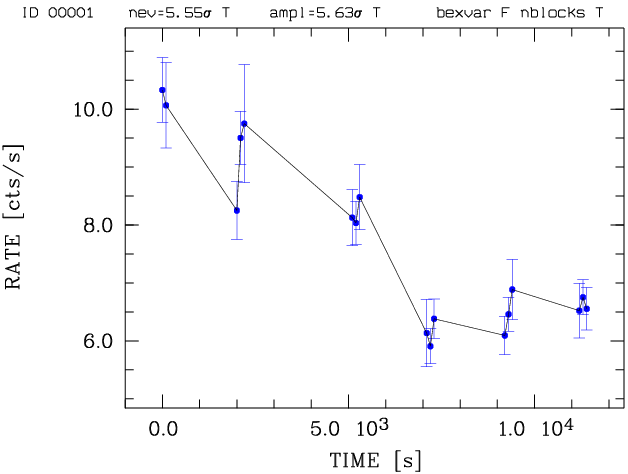}
    \includegraphics[height=4.7cm]{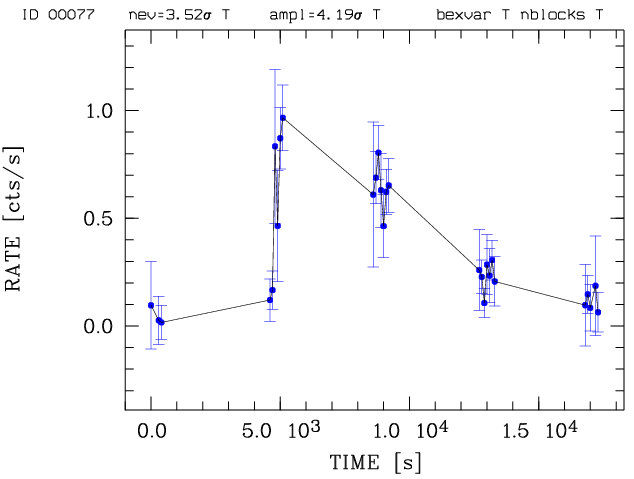}
    \includegraphics[height=4.7cm]{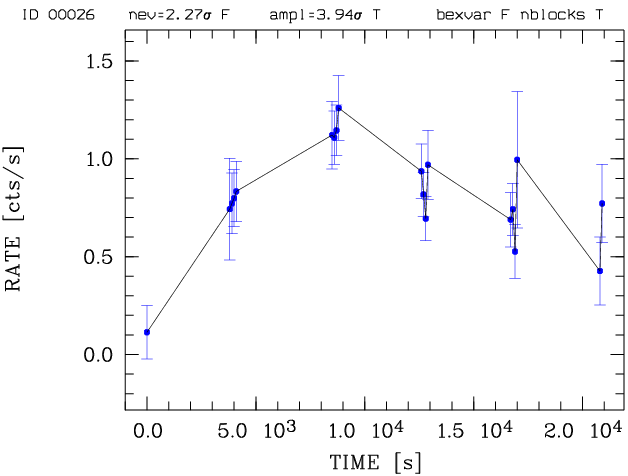}
    \includegraphics[height=4.7cm]{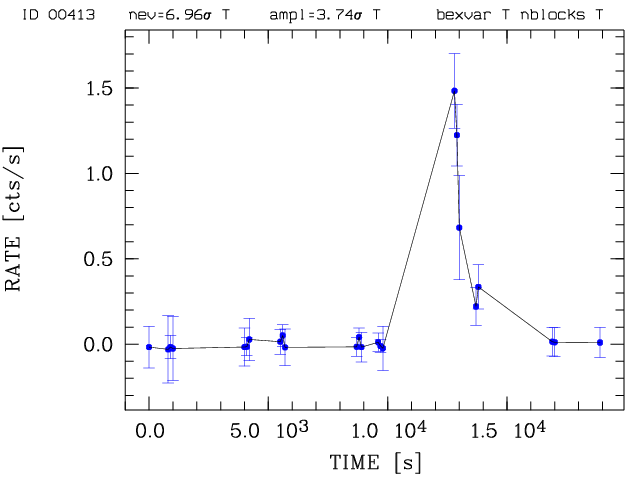}
    \includegraphics[height=4.7cm]{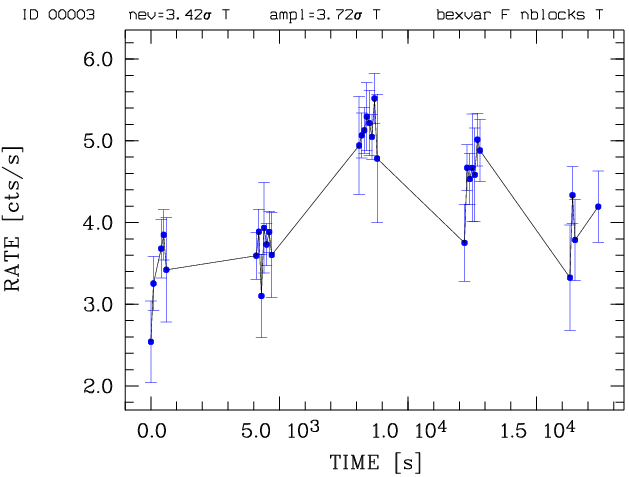}
    \includegraphics[height=4.7cm]{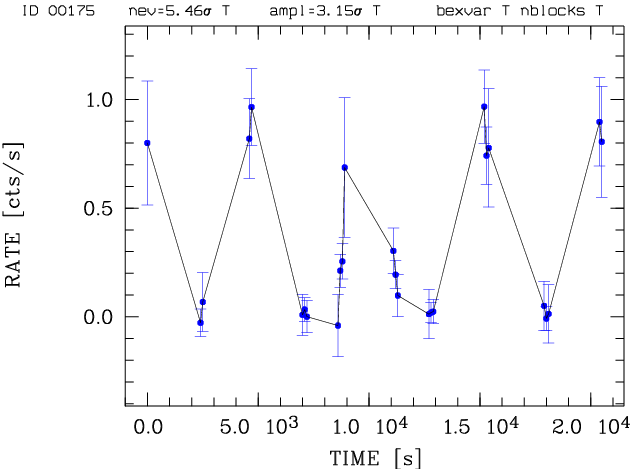}
    \includegraphics[height=4.7cm]{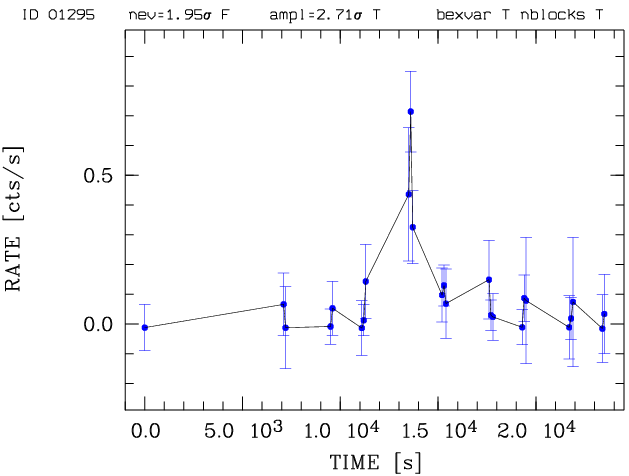}
    \includegraphics[height=4.7cm]{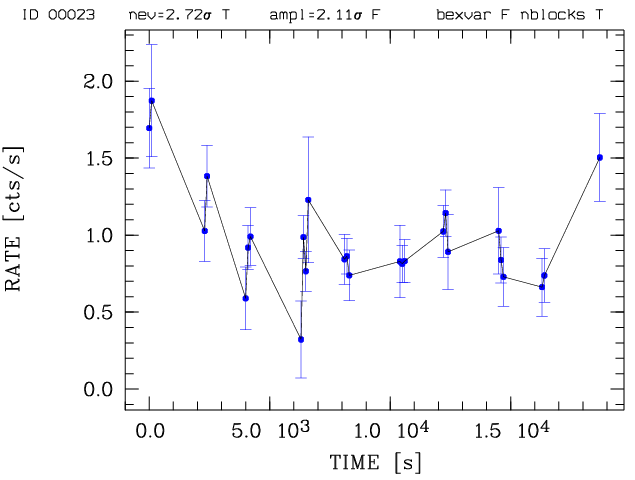}
    \caption{Examples for the 12 most variable eFEDS sources. The bin size is 100 seconds and the energy band is band 1. The objects are ordered from top left to right bottom in descending maximum amplitude $\sigma$ values. For the normalized excess variance and the maximum amplitude variability the values in units of $\sigma$ are given on top of each figures and the variability flags T for true and F for false are listed also for the Bayesian excess variance (bexvar) and the Bayesian block method (nblocks) in addition.
}
   \label{fig:extremevariab_gehrels}
\end{figure*}

For all eFEDS sources we have calculated the normalized excess variability, maximum amplitude variability, the Bayesian excess variability and the Bayesian blocks values and then applied the significance thresholds obtained by \cite{Buchner2021}. In the soft (0.2-2.3) keV band, we find 80 sources for which at least one of the four variability tests results in a significant variability detection. In
Fig.~\ref{fig:sampledist} we show their distribution in average count rate and ampl\_max; note that in the low count rate regime, only bexvar
identifies variable sources, while most sources identified by NEV and ampl\_max are also identified by other tests.

The NEV and ampl\_max methods provide reliable variability test results in the high count rate regime and for testing variability at the 100 second binning.
The Bayesian excess variance and the Bayesian block method select in addition variable sources at lower count rate statistics. 
To validate the significances especially in the low count rate regime, we have applied a validation method by comparing the observed counts with the simulated counts for a constant source. We consider a source as variable when at least in one time bin the number of observed counts is above 3 $\sigma$ compared to the constant count rate. These methods therefore test variability on longer time scales by merging data points between different pathes of sources through the 
FOV. Applying this 3 $\sigma$ count criterion, we find reliable variability for 65 out of the 80 selected sources in the (0.2-2.3) keV energy band. In 
Fig.~\ref{fig:3sigmacounts} we show the light curves and the cumulative count plots for source ID 2 in the high count rate regime and for source 837 as an example for variability in the low count rate regime.
In the hard (2.3-5.0 keV) band only source ID 2 is found as variable source based on the Bayesian excess variance test. This source is also a variable source in the soft band.  To give an impression of the obtained light curves, we show -- in Fig.~\ref{fig:extremevariab_gehrels} -- the 12 ``most'' variable sources found in our variability search.

\begin{table*}
\small
\caption{
List of the most variable eFEDS sources in the energy band (0.2-2.3) keV 
ordered in descending maximum amplitude values}.
\label{tab:ids}
\begin{tabular}{rlrrrrrrrllll}
\hline
  ID      & RA$^a$     &  DEC$^a$ & NEV$^b$& AMP$^c$&  GAIA g´$^d$& BP-RP$^e$& parallax$^f$& PM$^g$ &NW$^h$   & type detail        & X-ray$^i$\\
          &            &          &        &        &             &          &             &        &         &                    &         \\
2$^j$     & 138.72301  &  4.44351 & 20.95  & 15.42  &  7.65       & 1.30     & 55.93      & 77.9    & SG      & K/M binary         & 2RXS     \\
6$^j$     & 131.17500  &  0.73796 & 13.59  & 10.06  &  9.93       & 1.32     & 15.40      & 108.5   & SG      & K/M binary         & 2RXS     \\
95        & 143.16701  &  0.44491 &  7.74  & 7.25   & 17.13       & 2.54     &  3.25      & 18.6    & SG      & M2 star            & -        \\
5         & 139.16600  &  0.72883 &  3.69  & 7.23   &  6.66       & 0.56     & 14.60      & 92.0    & LG      & F/later type       & 2RXS     \\
1         & 144.25400  &  1.09597 &  5.43  & 5.63   & 15.97       & -        & -          & -       & SE      & Seyfert 1          & 2RXS     \\
77        & 143.78000  &  4.39564 & 3.37   & 4.19   & 18.42       & 3.72     &  9.71      & 185.2   & SG      & M5.5 star          & -        \\
26        & 130.92599  &  4.72626 & 2.29   & 3.94   & 12.65       & 1.37     &  1.03      & 22.7    & SG      & K4.5 star          & -        \\
413       & 132.57500  & -0.30916 &  6.93  & 3.74   & 17.02       & 3.40     & 16.93      & 123.1   & SG      & M5 star            & -        \\
3         & 134.07401  & -1.63477 & 3.41   & 3.72   & 17.21       & -        & -          & -       & SE      & Seyfert 1          & 2RXS/XMM \\
175       & 141.55901  &  1.09961 & 5.29   & 3.15   & 19.47       & 0.49     & 2.68       & 17.1    & SG      & CV                 & -      \\
1295      & 132.43300  &  2.64021 & 1.95   & 2.71   & 16.21       & 2.94     & 6.54       & 2.53    & SG      & M3 star            & -        \\
23$^k$    & 139.04300  &  1.88520 & 2.47   & 2.11   & 11.69       & 2.76     & 63.89      & 115.1   & SG      & M3 star            & 2RXS     \\
358       & 135.18800  &  0.02984 & 1.06   & 2.11   & 15.78       & 2.69     & 5.13       & 18.7    & SG      & M3 star            & -        \\
837       & 128.77391  &  3.03628 & -      & 2.05   & 17.36       & 3.26     & 5.873      & 18.3    & SG      & M5 star            & -        \\
226       & 131.05200  &  1.19534 & 2.85   & 2.04   & 12.56       & 1.18     & 2.48       & 23.1    & SG      & K3 star            & -        \\
837       & 128.77391  &  3.03629 & -      & 2.04   & 17.36       & 3.26     & 5.873      & 18.3    & SG      & M5 star            & -        \\
223       & 143.15965  &  1.15046 & -      & 2.03   & 19.40       & 0.71     & 0.07       & 7.0     & SG      & CV                 & -       \\
306       & 134.50668  & -2.66245 & -      & 1.92   & 20.02       & 0.96     & -0.25      & 0.7     & LE      &J085801.77-023945.9 & 2RXS     \\
766       & 131.18000  &  5.10186 & 2.51   & 1.89   & 12.52       & 1.26     & 6.61       & 22.5    & SG      & K3 star            & -        \\
4         & 130.10600  & 3.550909 & 0.67   & 1.88   & 17.76       &          &            &         & SE      & Seyfert 1                & 2RXS     \\ 
1402      & 133.32779  & -0.68684 & -      & 1.86   & 12.21       & 0.75     & 1.70       & 3.9     & SG      & GSC, F9.5          & -        \\
564       & 132.98309  &  4.06885 & -      & 1.83   & 18.76       & 3.24     & 4.00       & -       & SG      & -                  & -       \\
1770      & 138.05299  & 1.89884  & 0.14   & 1.71   & 16.74       & 2.88     & 3.34       & 21.9    & SG      & M4 star            & -        \\
154       & 139.23700  &  4.83086 & 3.86   & 1.60   & 14.13       & 1.11     & 1.79       & 33.1    & SG      & K2 star            & -        \\
284       & 137.47400  &  5.20347 & 2.45   & 1.53   & 8.07        & 1.19     & 38.27      & 38.2    & SG      & HD 78727           & 2RXS    \\
378       & 133.18800  & -1.61449 & 2.74   & 1.42   & 16.12       & 1.73     & 1.87       & 8.3     & SG      & K8 star            & XMM      \\
1800      & 128.43800  &  1.15426 & 1.92   & 1.40   & 18.76       & 3.33     & 6.68       & 68.9    & SG      & M5 star            & -        \\
5462      & 131.56800  &  1.63307 & -      & 1.34   & 19.27       & 1.99     & 0.99       & 6.32    & SG      & M0 star/multiple   & -        \\
25        & 140.42734  &  2.51763 & 1.94   & 1.29   & 11.18       & 0.94     & 5.99       & 34.2    & SG      & G9 star            & 2RXS     \\ 
642       & 132.67600  &  2.91918 & 1.91   & 1.17   & 14.38       & 1.73     & 4.76       & 19.3    & SG      & K8 star            & -        \\
2401      & 128.41542  & -0.28630 & -      & 1.03   & 14.28       & 1.08     & 0.47       & 6.18    & SG      & -                  & -       \\
681       & 130.03400  & -1.63099 & 1.23   & 1.03   & 14.76       & 2.57     & 10.87      & 41.7    & SG      & M3 star            & -        \\
1296      & 138.13300  & -0.79393 & 1.11   & 0.71   & 16.21       & 2.94     & 6.53       & -       & SG      & M4 star            & -        \\
\hline
\end{tabular}
\\
Notes:\\    
$^a$ Units in degrees. \\
$^b$ Normalized Excess variability in units of $\sigma$  \\
$^c$ Maximum amplitude variability in units of $\sigma$  \\
$^d$ GAIA G-band mean magnitude (Vega)   \\
$^e$ GAIA BP-RP colour  \\
$^f$ GAIA absolute stellar parallax in milli-second of arc   \\
$^g$ GAIA total proper motion in milli-sec of arc per year   \\
$^h$ NW = NWAY; classification based on \cite{Salvato2021}; SG=secure Galactic: LG=likely Galactic; SE=secure extragalactic; LE=likely extragalactic \\
$^i$ detection in 2RXS or XMM-Newton observations \\
$^j$ unresolved binary of a K-type and a M-type companion \\
$^k$ brightest M-dwarf detected in eFEDS \citep{Maggaudda2021}
\\
\\
\end{table*}

\subsection{The most variable eFEDS sources}

In Table~\ref{tab:ids} we list the most variable objects which have passed at least one of the significance threshold tests ordered by descending maximum amplitude variability in terms of $\sigma$.
We adopt the most likely counterpart identifications from \cite{Salvato2021}.
The vast majority of the counterparts to the highly variable X-ray sources show significant Gaia parallaxes or proper motions, indicating that they are secure Galactic objects (following the same classification scheme used by \cite{Salvato2021}). The spectral type of the stellar counterpart is also shown in the Table, based on publicly-available spectra. Four objects in this table are classified as extragalactic (based again on the \citet{Salvato2021} criteria) and we list the AGN classification type where available. 
Potential light leak false positives, with GAIA $g < 4.5$~mag that are detected exclusively in the soft band,  have been removed from the sample.
As apparent from Table~\ref{tab:ids}, the great majority of the variable sources are of galactic origin and are stars typically of spectral type K and M; this is in line with very similar findings in the study of variability in the ROSAT all-sky
survey reported by \cite{fuhrmeister2003}.  In the following we present more detailed investigations of individual sources and comparisons to related studies.

\subsection{Stellar flare events}

\begin{figure*}
    \includegraphics[width=\columnwidth]{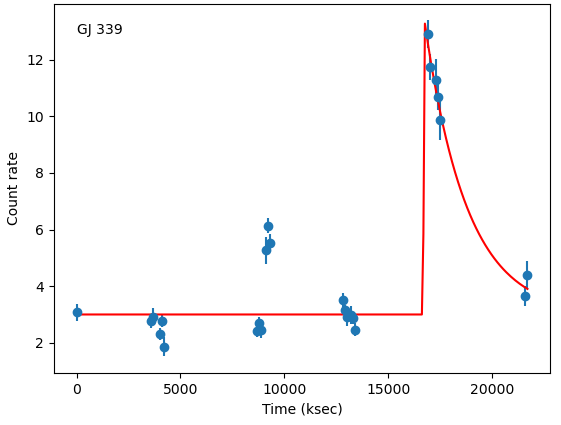}
    \includegraphics[width=\columnwidth]{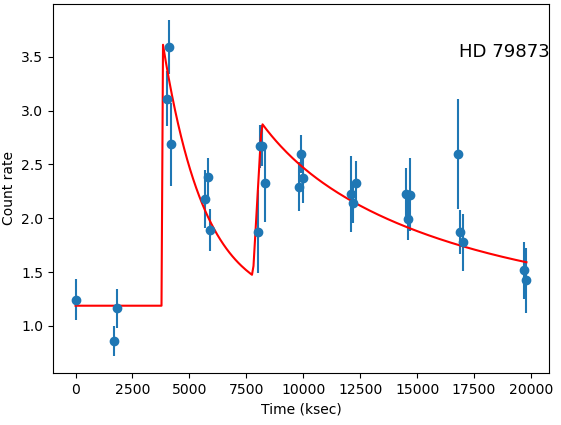}
    \includegraphics[width=\columnwidth]{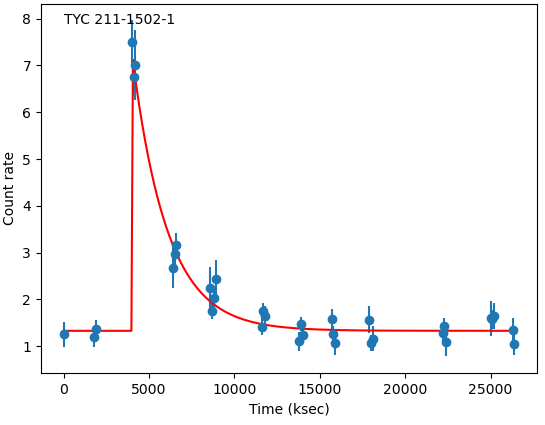}
    \includegraphics[width=\columnwidth]{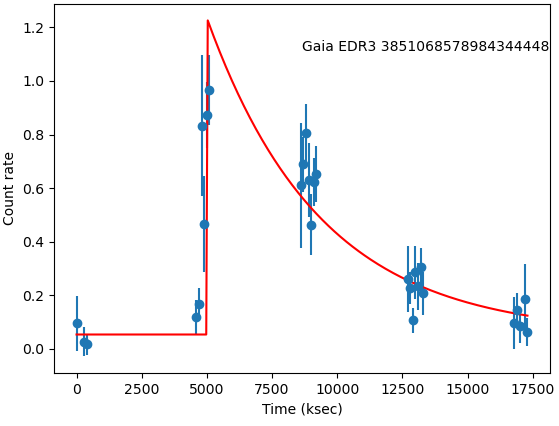}
    \includegraphics[width=\columnwidth]{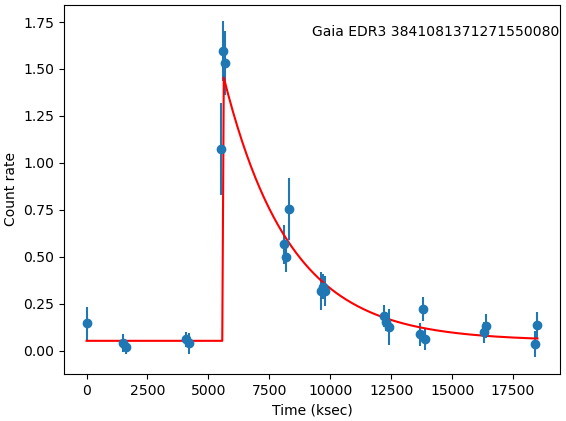}
    \includegraphics[width=\columnwidth]{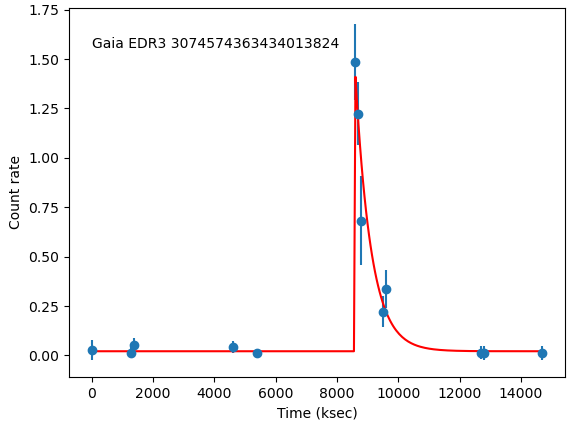}
    \caption{Fits to stellar flares for objects in Table 1 with source identifications 2, 5, 6, 77, 95, 413 (from top left to right bottom).
    }
    \label{fig:flarestars}
\end{figure*}


\begin{table*}[th]
\begin{center}
\caption{\label{tab2flares}eFEDS sample of stellar flares}
\begin{tabular}{r l l l l l c c}
\hline
ID        & L$_{bol}$   & log L$_{X,quiet}$ &$t_{decay}$  & log L$_{X,peak}$ & log E$_{tot}$ & log L$_{X,quiet}$/L$_{bol}$ & log L$_{X,flare}$/L$_{bol}$\\
          & (solar units)& (erg/s) & (sec)  & (erg/s)     & (erg)           & & \\
\hline
2$^1$         & 0.22      & 29.0   & 2000        & 29.6               & 32.9        & -3.9     & -3.40\\
5$^2$     & 3.18      & 29.8   & 1800        & 30.1               & 33.4        &-4.3      & -4.0\\
5$^3$     & 3.18      & 29.8   & 9000        & 29.9               & 33.8        &-4.3      & -4.2\\
6$^4$     & 0.22      & 29.8   & 2000        & 30.4               & 33.7        & -3.1     & -2.5\\
77        & 0.0014    & 28.8   & 4300        & 30.1               & 33.8        & -1.9     & -0.6\\
95        & 0.0156    & 29.7   & 2700        & 31.2               & 34.6        & -2.0     & -0.6\\
413       & 0.0025    & 27.9   & 500         & 29.7               & 32.4        & -3.1     & -1.25\\
\hline
\end{tabular}
\\
\end{center}
Notes:\\
$^1$ 
Flare energetics are determined for the event at around 17.000 seconds. The decay time for the potential flare event at about 9000 seconds cannot be determined uniquely. We therefore do not apply a second flare fit to the source.
\\
$^2$ Numbers refer to the first flare event. \\
$^3$ Numbers refer to the second flare event with a longer decay time.  Note that ID00005 is -- 
according to Gaia -- a visual binary with a separation of 2.4"; we have assumed that the flare occurred
on the hotter star.\\
$^4$ ID00006 is -- according to Gaia -- a visual binary with a separation of 3.2"; we have assumed that the flare occurred
on the hotter star.\\
\end{table*}

Because of the scanning character of eROSITA's eFEDS observations, the detected X-ray sources
are not continuously covered.  For stellar flares this creates a 
problem,
since important parameters like the time of flare onset, time and amplitude of
flare peak and flare decay may not or may only be partially sampled. Clearly, if
there is only one high data point in a light curve, we can note that a flare most likely
occurred, but little or nothing can be inferred about the flare energetics.
By visual inspection we identified six stellar flares (cf., Tab.~\ref{tab2flares}), with
sufficient X-ray coverage to allow a meaningful flare characterization.

To describe the flares, we use an empirical {\it ad hoc} model, assuming
a constant coronal background $B$, interpreted as the quiescent X-ray luminosity,
a linear increase from the time of flare onset
($t_{onset}$) to the time of flare peak ($t_{peak}$) with some flare amplitude
$A$, followed by an exponential decay with some characteristic time scale
$t_{decay}$, i.e.,  the model is thus defined by five adjustable parameters.
Obviously, such a ``model'' is quite simplistic, yet many (albeit not all) stellar flares
can be reasonably described by such an ansatz.  The X-ray light curves together with our fits are
presented in Fig.~\ref{fig:flarestars}; note that formally all our fits are statistically
acceptable.

Even then there may be remaining ambiguities depending on how the light curve was sampled.  
The time of flare onset is usually not observed, and, given the scanning character of our observations,
the time of flare peak and hence the peak amplitude are often also not observed.
While the former is not that relevant for the overall flare energetics (only a
small fraction of the X-ray energy is radiated away in the rise phase), the total
energy in the decay phase is directly proportional to the -- typically unobserved --
flare amplitude.  We therefore attempted to construct ``minimum energy'' models,
i.e., models that would yield the lowest peak X-ray luminosities and lowest
X-ray energies; our numbers should therefore be considered lower limits.

In Tab.~\ref{tab2flares} we list the sources analyzed in detail and the
derived flare parameters. We specifically provide the source ID, the inferred
bolometric luminosities (in solar units), the measured quiescent
X-ray luminosities, the derived decay time scale, observed peak X-ray luminosities
and total X-ray energies as well as the respective $L_{X}$/$L_{bol}$ ratios w.r.t. the
quiescent and peak X-ray luminosities; the X-ray data are derived in this paper, the bolometric
luminosities were computed from the tables provided by \cite{pecaut2013}.
Concerning Table~\ref{tab2flares}
a few words are in order:  Two of the flaring stars are actually visual binaries that cannot be separated by
eROSITA and hence the flare site is unknown. Source ID2 (= TYC 211-1502-1) consists of
a K-type star and a M-type star at the same distance but separated by 3.2", source ID5
(= HD~79873) consists of a F-type and a later-type companion at an angular distance of 2.4", the rest of the host stars appear to be single using GAIA EDR3.  Inspecting the decay
time scales listed in Tab.~\ref{tab2flares}, one notes that in most cases the decay times are short (500 sec -  2000 sec). During the eFEDS observations sources were typically exposed for 400 seconds, so that usually such decay times are reasonably well defined.  Only the flare(s) observed on HD~79873 lasted longer;
the long-duration flare on HD~79873 appears to have consisted out of two
individual events with much longer decay time scales.

Of special interest are the flares observed on ID77 (Gaia EDR3 3851068578984344448), ID95 (Gaia EDR3 3841081371271550080) and ID413 (Gaia EDR3 3074574363434013824). All of these flares are observed on anonymous stars, the Gaia ERD3 IDs of which are listed,  and all of these three stars are very red and very faint (cf., Gaia g magnitudes and BP-RP colors listed in Tab.~\ref{tab:ids}). The eROSITA and Gaia positions agree very well, and at least no other Gaia sources are located in the vicinity.  The flares observed by eROSITA -- which do look like textbook stellar flares -- therefore make our identifications unequivocal.   ID95 appears to be outstanding; according to Gaia it is located at a distance of 308~pc with an uncertainty of about 10~pc, consequently its quiescent and flaring X-ray luminosity are quite large.

Both for ID77 and ID95 the ratio between the observed peak X-ray luminosity and the quiescent
bolometric luminosity approaches unity.  Observations of late-type dwarfs with {\it Kepler} and TESS have demonstrated
that flares on late-type stars can produce significant enhancements even in their broad band
optical light curves, i.e., in the case of TESS in the band between 6000 \AA \, and 10000 \AA; note that ground-based flare observations are typically carried out in the
U or B bands.  The most extreme event in the study of TESS observed optical flares reported
by \cite{guenther2020} occurred on the M4.5 star Gaia EDR3 5495481200071568640/2MASS J06270005-5622041, which increased its
optical flux (in the TESS band) by a factor of 16.1 with a total energy release of 10$^{34.7}$ erg;
with its color of BP-RP = 2.98 it is quite similar to the stars considered here.
\cite{schmitt2019} analyzed TESS and XMM-Newton flare observations of the young
active star AB~Dor (spectral type K0) and demonstrated that the optical output of such flares is likely on 
the order of the X-ray output and possibly larger by up to one order of magnitude.  In that context
we note that the total energy output of our most extreme X-ray event on ID95 coincides 
nicely with that obtained for Gaia EDR3 5495481200071568640 in the optical.

What maximal ratios between the observed peak X-ray luminosity and the quiescent
bolometric luminosity approaches are then reasonable?
Consider then an M dwarf with $T_{eff} = $ 3000~K and a peak flare X-ray flux of 10$^{31}$ erg/s,
and assume that during the flare $T_{eff,flare}$ = 12000~K, which is a typical chromospheric
temperature, and $L_{opt,flare}$ = $L_{X,flare}$ = 10$^{31}$ erg/s.   Under these assumptions
a flare area of less than one percent of the stellar surface would be sufficient to lead to $L_{star,flare}$ =
$L_{star,quiet}$, and coverage fractions of a few percent would be required to account for extreme cases
like Gaia EDR3 5495481200071568640/2MASS J06270005-5622041.  Incidentally Gaia EDR3 5495481200071568640/2MASS J06270005-5622041
is seen as a saturated X-ray source in eRASS1 with $L_{X,quiet}$/L$_{bol,quiet}$ = -2.45, and likely to produce
X-ray flares similar to those described here.

\subsection{Remarks on the AGN sub-sample}

Of the discovered intrinsically variable sources, the majority is classified as "Secure Galactic" or "Likely Galactic" according to \cite{Salvato2021}. Only 18 per cent are of extragalactic origin. 
Three objects are classified as Seyfert 1 galaxies (Mrk 707,  2MASX J08561784-0138073, and 2MASS J08402550+0333018). All of these objects have been detected in the 2RXS catalogue \citep{2016Boller}.
Four objects are classified as QSOs (SDSS J091000.01+034429.1, [VV2006] J085141.5+011930, [VV2006] J091637.5+004734, and  [VV2006] J092346.8+011142). Only the last object listed has an 3XMM X-ray detection. 
Three objects are classified as galaxies (SDSS J090905.13+010929.6, SDSSCGB 12543.2, and TXS 0929+017). None of these sources have been detected with previous X-ray missions. 
Two objects are identified as Brightest galaxy in a Cluster (BCG), SDSS J084150.93-001540.7 and SDSS J083125.18+034333.1, detected at X-rays for the first time with eROSITA.

\section{Comparison with 2RXS}

\begin{figure*}
\includegraphics[width=\columnwidth]{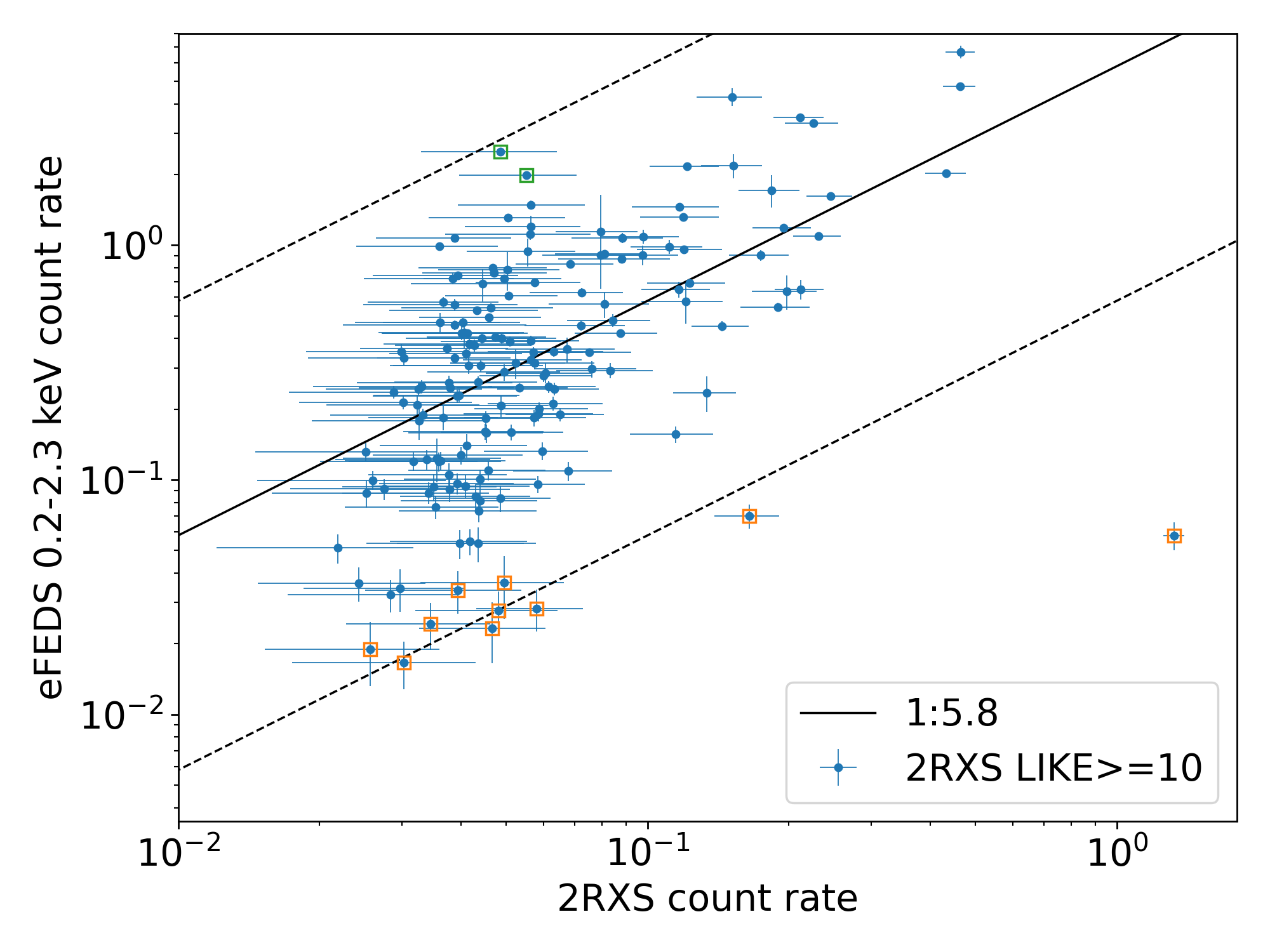}
\includegraphics[width=\columnwidth]{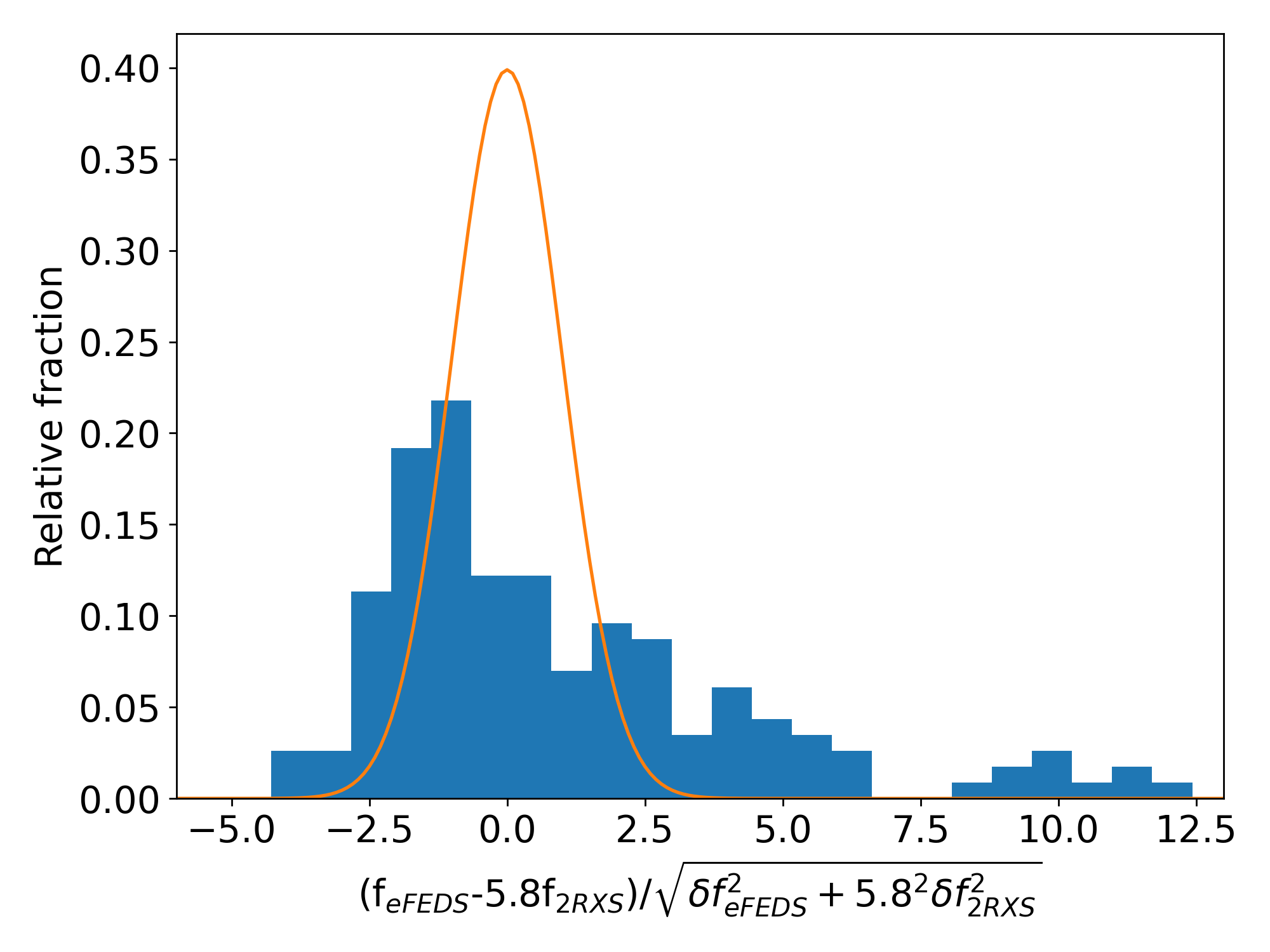}
    \caption{Left panel: 2RXS versus eFEDS count rates for the common sources detected in both catalogs. Only sources with 2RXS detection likelihoods  {\texttt {EXI\_ML$>$10}} and eFEDS detection  likelihoods {\texttt {DET\_LIKE$>$10}} have been cross-matched between both surveys to avoid that spurious detections bias the data analysis.
    Blue data points indicate sources with variability factors below 6 and the orange data points show sources with variability factors above 6.
    The black solid line corresponds to a median ratio of 5.8 and the dashed lines indicate offsets from the solid line by a factor of 10. 
    Right panel: Distribution of the flux difference between the eFEDS and 2RXS sources with detection likelihoods in both catalogs above 10 with respect to the median ratio line (the black solid line in the upper panel) in terms of flux uncertainty. The orange line displays the standard normal distribution (unity scatter, zero mean).}
    \label{fig:2RXSeFEDS}
\end{figure*}

We cross-correlated the 2RXS catalogue \citep[][]{2016Boller} with the eFEDS main catalog by searching for the nearest eFEDS source within 30\arcsec of any 2RXS source. 
322 2RXS sources are matched to 359 eFEDS sources in this way.
Fig.\ref{fig:2RXSeFEDS} (left panel) compares the eFEDS 0.2-2.3 keV count rates and the 2RXS count rates for the nearest matched sources for detection likelihood values above 10 for both surveys. 
A conversion factor between these two count rates is measured as the median of the ratio between 2RXS and eFEDS sources. 
There are 10 sources with a variability factors greater than 6 compared to normalized mean ratio between 2RXS and eFEDS sources which are marked as orange squares.
Two sources are at least 6 times brighter during the eFEDS observations compared to 2RXS marked as green squares (c.f. Table~\ref{tab:factor6_2RXS_eFEDS}).
10 sources are below and another two sources are above the median line.

\begin{table*}[!hth]
\begin{center}
\begin{tabular}
{ l                      r          l             l                l               r        r         l       r            r                                 }
\hline
IAU NAME                 & rate$^1$  & err$^1$   &   RA 	    &    DEC          &   ID   &   rate$^2$  &  err$^2$   & factor      & NW$^3$                  \\   
(1)                      & (2)       & (3)       &   (4)        &    (5)          &   (6)  &   (7)       &  (8)       & (9)         & (10)                  \\
\hline
2RXS J084717.7+030050    &   4.81    & 1.61      & 08h47m17.72s & +03d00m50.0s    &  7530  &   2.70      & 0.56       &  10.3       &  SG     \\ 
2RXS J084815.5+044537    &   4.95    & 1.68      & 08h48m15.51s & +04d45m37.5s    & 14621  &   3.60      & 1.10       &   8.0       &  SE     \\ 
2RXS J091407.5$-$011049  &   5.79    & 1.48      & 09h14m07.53s & $-$01d10m49.7s  &  7049  &   2.82      & 0.57       &  11.9       &  LE     \\ 
2RXS J091421.8+021914    &  16.48    & 2.59      & 09h14m21.88s & +02d19m14.8s    &  1089  &   7.01      & 0.81       &  13.6       &  SG     \\ 
2RXS J084104.1+032119    & 132.38    & 6.73      & 08h41m04.10s & +03d21m19.2s    &  2558  &   5.79      & 0.77       & 132.7       &  SG     \\ 
2RXS J085035.3$-$003432  &   4.66    & 1.40      & 08h50m35.33s & $-$00d34m32.8s  & 13437  &   2.32      & 0.67       &  11.6       &  LE     \\ 
2RXS J083539.3$-$014129  &   3.02    & 1.28      & 08h35m39.32s & $-$01d41m29.8s  &  9141  &   1.66      & 0.38       &  10.5       &  SG     \\ 
2RXS J085031.1+015658    &   2.57    & 1.04      & 08h50m31.18s & +01d56m58.2s    & 13004  &   1.89      & 0.58       &   7.9       &  LG     \\ 
2RXS J091849.6+021507    &   3.94    & 1.44      & 09h18m49.64s & +02d15m07.4s    &  4104  &   3.38      & 0.69       &   6.8       &  SE     \\ 
2RXS J091254.6$-$013848  &   3.45    & 1.17      & 09h12m54.68s & $-$01d38m48.4s  &  4219  &   2.42      & 0.54       &   8.3       &  LE     \\ 
2RXS J085755.3+031059    &   4.85    & 1.56      & 08h57m55.36s & +03d10m59.7s    &   108  & 250.00      & 8.88       &  0.11       &  -      \\ 
2RXS J090320.6+045758    &   5.51    & 1.54      & 09h03m20.64s & +04d57m58.1s    &    35  & 199.10      & 7.67       &  0.16       &  SE     \\ 
\hline
\end{tabular}
\caption{eFEDS - 2RXS sources with variability factors greater than 6:
The columns are 
(1):   2RXS X-ray source identification number; 
(2,3): 2RXS count rates and corresponding errors in the 0.1-2.4 keV energy band;
(4,5): 2RXS X-ray sky coordinates, right ascension and declination in J2000;
(6):   eFEDS source identification number;
(7,8): eFEDS count rates and corresponding errors in the 0.2-2.3 keV energy band;
(9):   factor of variability between the 2RXS (medium ratio factor of 5.8 applied) and the eFEDS count rates;
(10):  object classification type details
}
\label{tab:factor6_2RXS_eFEDS}
\end{center}
Notes:\\
$^1$ 2RXS count rates and corresponding errors in units of $\rm 10^{-2}\ cts/s$. The medium ratio factor of 5.8 has to be applied for comparison with the eFEDS count rates \\
$^2$ eFEDS count rates and corresponding errors in units of $\rm 10^{-2}\ cts/s$  \\
$^3$ NW = NWAY; classification based on \cite{Salvato2021} \\ 
\end{table*}

\section{Comparison with the XMM-ATLAS field}

The eFEDS field overlaps with one of the wide-area and shallow surveys carried out by XMM-Newton, i.e., the XMM-ATLAS survey 
\citep{Ranalli15}. The XMM observations were obtained in May 2013 and
cover a total area of about $\rm 6\,deg^2$. Comparing this data set with the eFEDS observations provides the opportunity to explore the X-ray variability over a 6.5\,yr timescale. This section searches for X-ray sources which show evidence for significant variations in their X-ray flux between the eFEDS and the XMM-ATLAS epochs. The comparison is not limited to sources detected in both observations but also uses upper limits in the case of sources detected by one telescope but not the other.

A custom reduction of the XMM-ATLAS survey field is used based on the methods described by \cite{Georgakakis_Nandra11}. The advantage of using a custom analysis rather than the publicly available XMM-ATLAS catalogue, is a better control over the X-ray sensitivity, which is important for the calculation of upper limits. The relevant XMM-Newton observations (with identification numbers 0725290101, 0725300101, 0725310101) were reduced using the XMM-Newton Science Analysis System (SAS) version 18. Sources are detected independently in 3 energy intervals, 0.5-2,  2-8  or 0.5-8\,keV  with a Poisson  false  detection threshold  of $\rm < 4 \times 10^{-6}$. A total of 987 sources were detected in the 0.5-2\,keV band to the threshold above. The calculation of fluxes is based on aperture photometry 
using the Bayesian methodology described by \cite{Laird09} and \cite{Georgakakis_Nandra11}. The advantage of this approach is that it accounts for the Eddington bias in the determination of fluxes, which is expected to become important in the low-count regime of these X-ray imaging observations. Assuming that $N$ is the total number of counts extracted within an aperture of radius $R$, and $B$ is the corresponding background level, then the probability of a source with flux $f_X$ is given by the Poisson formula

\begin{equation}\label{eq:age:Pf}
P(f_X | N)  = \frac{e^{-\lambda}\cdot \lambda^N}{N!}\,\pi(f_X)
\end{equation}

\noindent where $\pi(f_X)$, reflects the prior knowledge on the distribution of source fluxes i.e. the differential X-ray source number counts described by a double power-law \citep[e.g.][]{Georgakakis08}. It is this term that accounts for the Eddington bias. The quantity $\lambda$ is the expected number of photons in the extraction cell for a source with flux $f_X$

\begin{equation}\label{eq:age:lambda}
\lambda = f_X\cdot EEF\,\sum_{i=1}^{3} ECF_{i}\cdot t_{i} + B,
\end{equation}

\noindent where $t_{i}$ is the exposure time, $ECF_{i}$ is the energy flux to photon flux conversion factor and depends on the spectral shape of the source. The summation in equation above is for the 3 EPIC detectors, PN, MOS1 and MOS2. The calculation above assumes that the radius $R$ of the extraction aperture corresponds to a fixed Encircled Energy Fraction (EEF) of the XMM-Newton PSF. In our analysis the value of the EEF is set to 80\%. Equation \ref{eq:age:Pf} can be numerically integrated \citep[see][]{Kraft91} to determine the most likely value of $f_X$ and the corresponding uncertainties at a given confidence level. In the case of non-detections the same equation can be integrated to determine upper limits. Fluxes for the XMM-ATLAS sources are estimated in the 0.5-2\,keV spectral band assuming a power-law spectral model with $\Gamma=1.4$ that is absorbed by a Galactic column density of $\log N_H/\rm cm^{-2}=20.3$.

The main catalogue of the eFEDS field consists of sources detected in a single spectral band, 0.6-2.3\,keV. This was chosen because the sensitivity of eROSITA peaks in this energy interval. Forced photometry in other spectral bands, including the 0.5--2\,keV, is also available at the positions of the X-ray sources of the main eFEDS catalogue. Two independent approaches were adopted for the X-ray photometry. The first fits a model of the eROSITA Point Spread Function (PSF) to the 2-dimensional distribution of X-ray photons at a given position ({\sc{ermldet}} task of eSASS). The second extracts counts within an aperture with size that corresponds to a fixed Encircled Energy Fraction ({\sc{ apetool}} task of eSASS). Here we use the aperture photometry products (total counts, background level, exposure time) to determine the source fluxes in the 0.5-2\,keV band using the Bayesian methodology outlined above. The same spectral model as in the case of the XMM-ATLAS analysis is adopted.  The aperture radius corresponds to the 60\% EEF. 
In the rest of the analysis we consider a total of 985 eFEDS sources that lie within the XMM-ATLAS footprint and with a detection likelihood in the 0.6--2.3\,keV band $DET\_LIKE>10$. The latter threshold is to avoid spurious detections.

\begin{figure}
\includegraphics[width=\columnwidth]{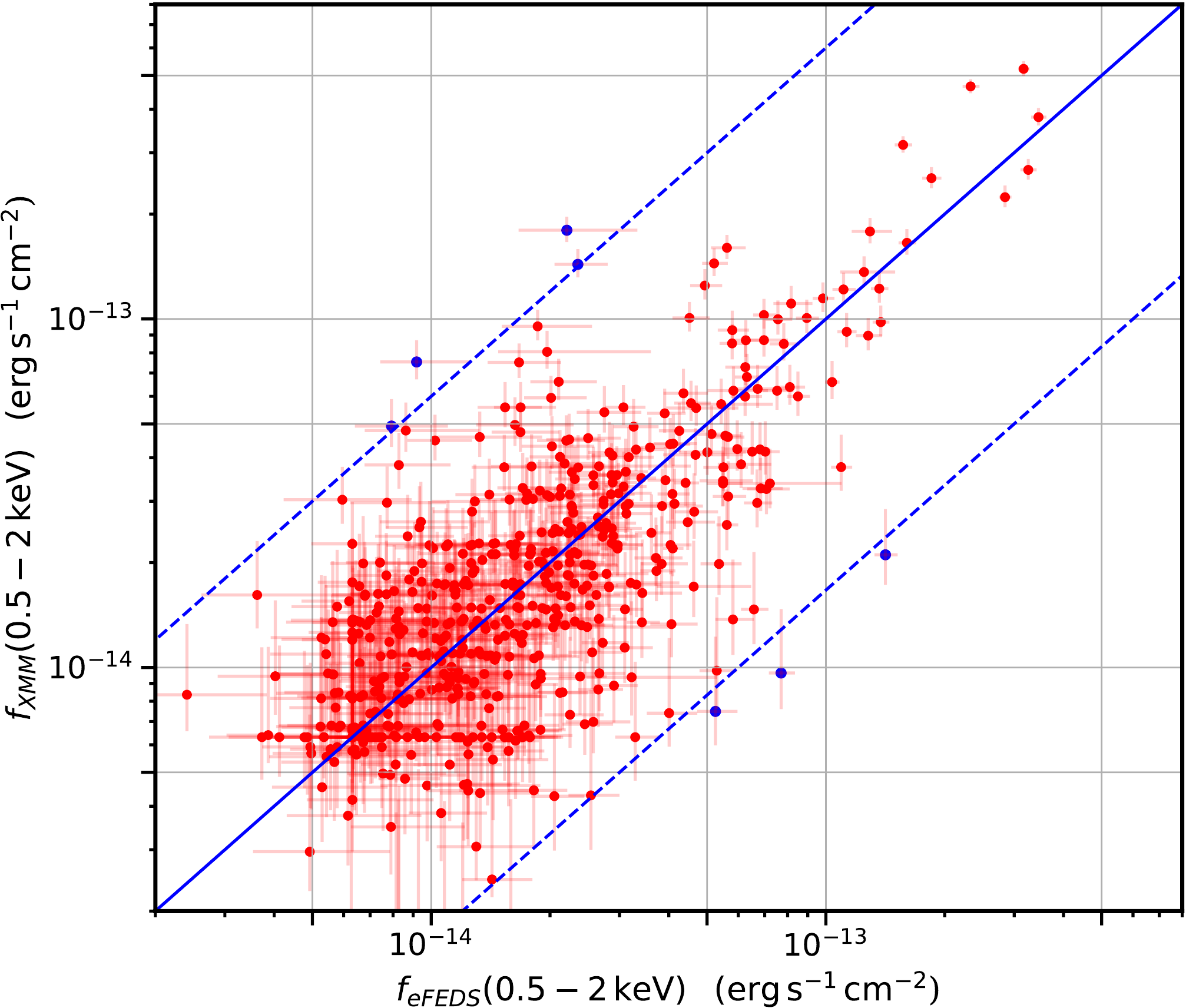}
\includegraphics[width=\columnwidth]{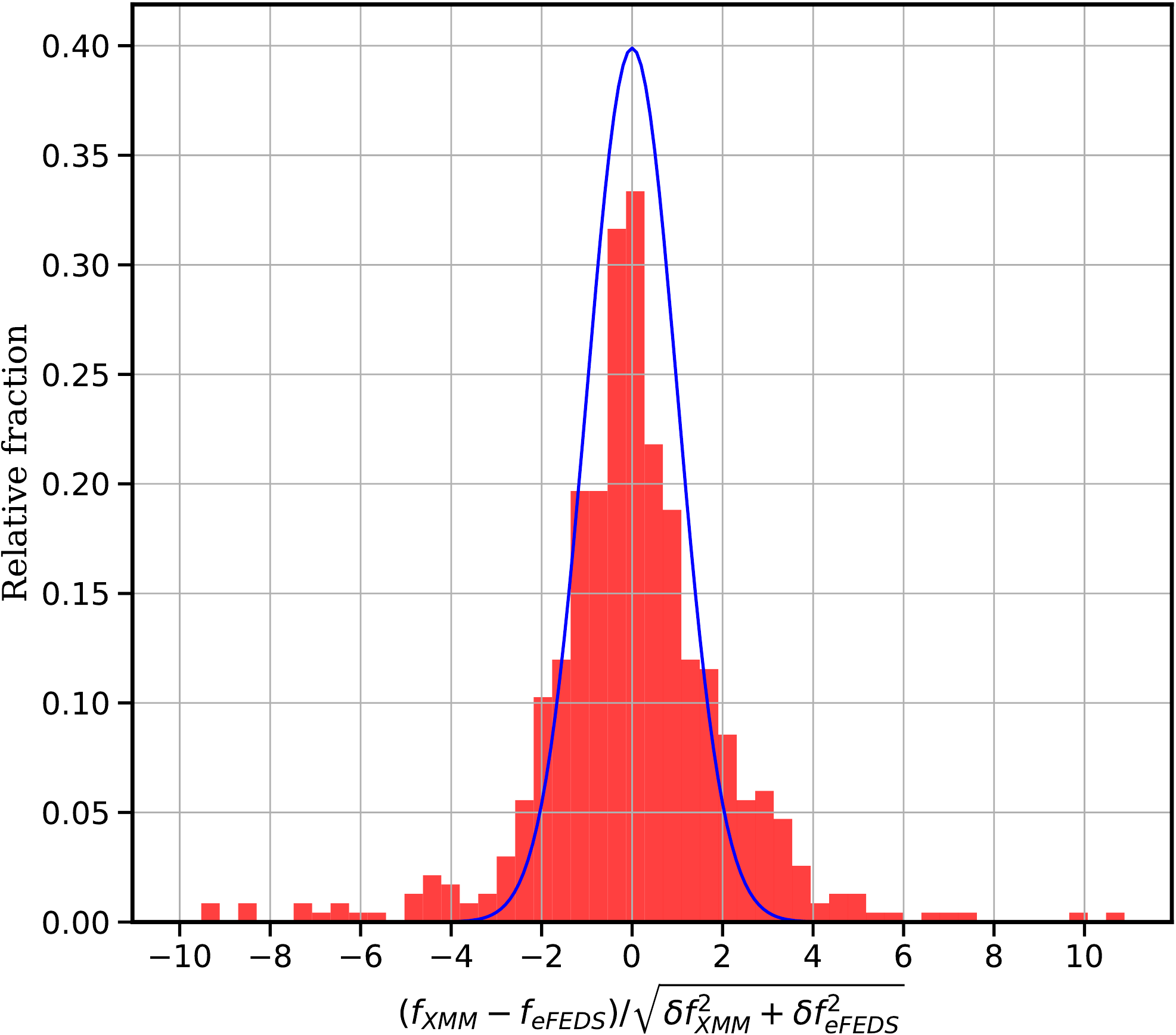}
\caption{The top panel shows the XMM-ATLAS versus eFEDS flux in the 0.5-2\,keV band. The red dots are the common sources between the two surveys. The error bars correspond to the 68 per cent uncertainty. The blue solid line shows the one-to-one relation. The blue dotted lines correspond to a flux ratio of 6. Sources with fluxes that differ by more than this factor in the two surveys are shown with blue colour. Bottom panel: The red histogram shows the distribution of the flux difference of X-ray sources between the  XMM-ATLAS and eFEDS observations normalised by the corresponding flux uncertainties added in quadrature. The blue line shows a Gaussian distribution with mean of zero and scatter of unity. The histogram of the normalised flux difference has more power in the wings compared to the normal distribution, which we attribute to variability.}
    \label{fig:age-ero-atlas-det}
\end{figure}

The first step is to match the eFEDS and XMM-ATLAS sources using a search radius of 15\,arcsec. For the sky density of the XMM and ATLAS sources this threshold corresponds to $\ll 1$
spurious associations. Fig.~\ref{fig:age-ero-atlas-det} compares the 0.5-2\,keV fluxes of the 616 common sources in the two surveys. The error bars correspond to the 68\% confidence interval. The data points in this figure scatter around the one-to-one relation without evidence for strong systematic offsets. The scatter increases toward fainter fluxes because of the larger shot noise. This is further explored in the bottom panel of Fig.~\ref{fig:age-ero-atlas-det}, which plots the histogram of the flux difference between eFEDS and XMM-ATLAS normalised to the flux errors added in quadrature. There are however, sources that show significant variations in their flux between the two epochs. We attribute such differences to the intrinsic variability of accretion events and highlight the most extreme examples by selecting sources with flux variations of at least a factor of 6. These sources are highlighted in Fig.~\ref{fig:age-ero-atlas-det} and are listed in Table \ref{tab:age-ero-atlas-det}. Four of these seven sources are associated with spectroscopically confirmed AGN at $z<1$.

\begin{figure}
\includegraphics[width=\columnwidth]{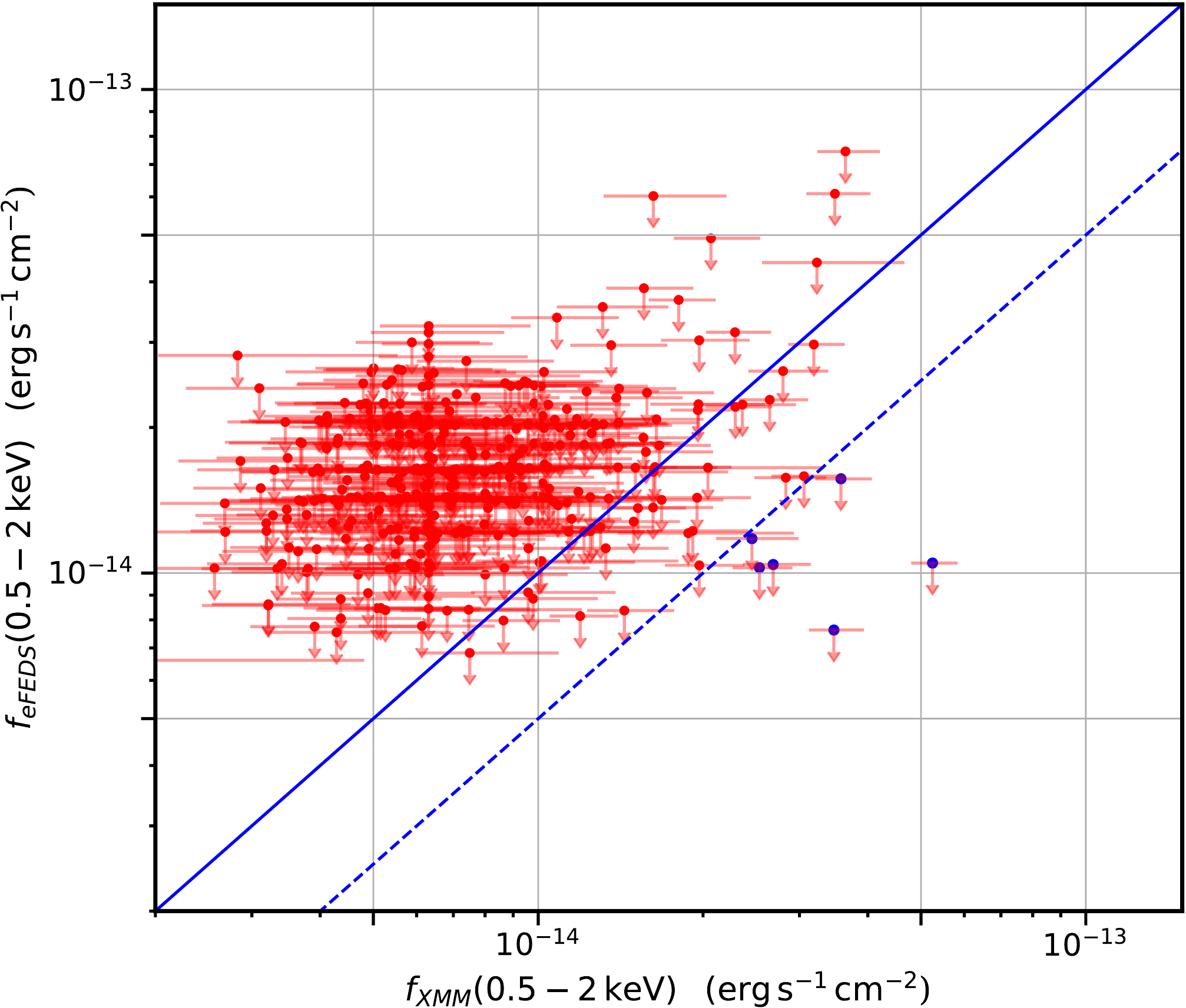}
   \caption{XMM-ATLAS 0.5-2\,keV flux versus eFEDS flux upper-limit ($3\sigma$ confidence) in the 0.5-2\,keV band. Each data point in this plot corresponds to an XMM-ATLAS source detection without counterpart in the eFEDS source catalogue. For these sources the $3\sigma$ flux upper limits are estimated from the eFEDS observations and plotted on the vertical axis. The blue dashed line indicates the one--to--one relation. The blue dotted line corresponds to an offset by a factor of two in flux relative to the one-to-one relation, i.e. sources with eFEDS $3\sigma$ flux upper limits that are two times fainter than the XMM-ATLAS source flux. Sources below that line are highlighted by blue colour. They have been visually inspected to exclude XMM ATLAS spurious sources or sources close to edge of the field of view. The resulting sample following the visual inspection is listed in Table \ref{tab:age:EFEDSUPPER}.}
    \label{fig:age:EFEDSUPPER}
\end{figure}

\begin{figure}
    \includegraphics[width=\columnwidth]{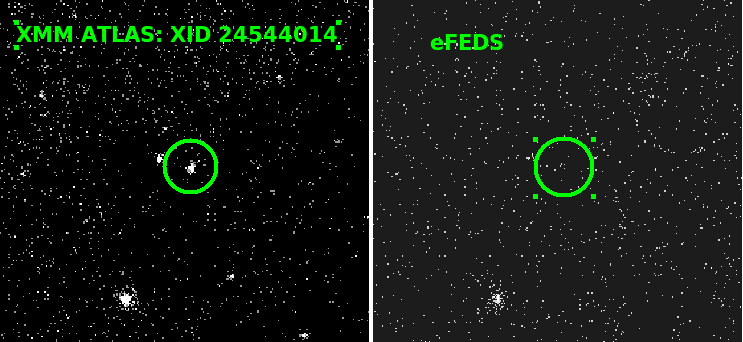}
    \caption{X-ray post-stamp image of a source detected in the XMM-ATLAS survey that has no counterpart in the eFEDS and a ratio between the XMM 0.5-2keV flux and the eFEDS 0.5-2.0 keV $3\sigma$ upper limit $>2$. The green circles mark the position of the source in the two observations and have 30\,arcsec radius.}
    \label{fig:age:EFEDSUPPER_EXAMPLE}
\end{figure}

The analysis above is based on the comparison of the X-ray fluxes of sources detected in both the XMM-ATLAS and eFEDS fields. It will miss variability events that reduce a source's flux below the formal detection limit at either the XMM or the eROSITA epoch. Examples of such variable sources are identified by looking into the XMM-ATLAS detections with no counterparts in the eFEDS survey and vice versa. The flux of a source at the detection epoch is compared with the $3\sigma$ upper limit on its flux estimated for the observation on which it lies below the detection threshold. This comparison can differentiate true variability events from non-detections because of either low effective exposure times (e.g. shallow observations or large off-axis angles) or high background levels. The estimation of upper limits uses aperture photometry and follows the methods described by \citet{Ruiz21} based on the work of \citet[][]{Kraft91}. The probability density function of Equation \ref{eq:age:Pf} is integrated with respect to $f_X$ to determine the corresponding cumulative distribution as a function of this parameter.  The upper limit, $UL$, at an arbitrary confidence interval $CL$ is then defined as the flux value at which the cumulative probability equals $CL$

\begin{equation}\label{eq:age:confidence}
C\cdot \int_{0}^{UL} P(f_X | N) \,df_X = CL,
\end{equation}

\noindent where $C$ is a normalisation constant. In the calculation of upper limits we assume a flat prior $\pi(f_X)$ in Equation \ref{eq:age:Pf} and a confidence level $CL=99.87$\% that the true flux lies below the respective upper limit. This probability corresponds to the one-sided $3\sigma$ limit of a Gaussian distribution. First we consider XMM-ATLAS sources with no eFEDS counterparts within 30\,arcsec. Using an aperture with radius of 60\% EEF we determine the eFEDS counts, background level and exposure time at the positions of the XMM-ATLAS detections. These are plugged into Equations \ref{eq:age:lambda} and \ref{eq:age:confidence} to determine the eROSITA epoch $3\sigma$ upper limits on the sources' fluxes.  The comparison between the eROSITA upper limits and the XMM-ATLAS fluxes is shown in 
Fig.~\ref{fig:age:EFEDSUPPER}.  Table \ref{tab:age:EFEDSUPPER} presents the selected sources with flux ratio between the eFEDS and XMM epochs $> 2$. 
As a demonstration Fig.~\ref{fig:age:EFEDSUPPER_EXAMPLE} shows the XMM-ATLAS and eROSITA/eFEDS X-ray post-stamp images of one of the sources presented in Table  \ref{tab:age:EFEDSUPPER}.

\begin{figure}
\includegraphics[width=\columnwidth]{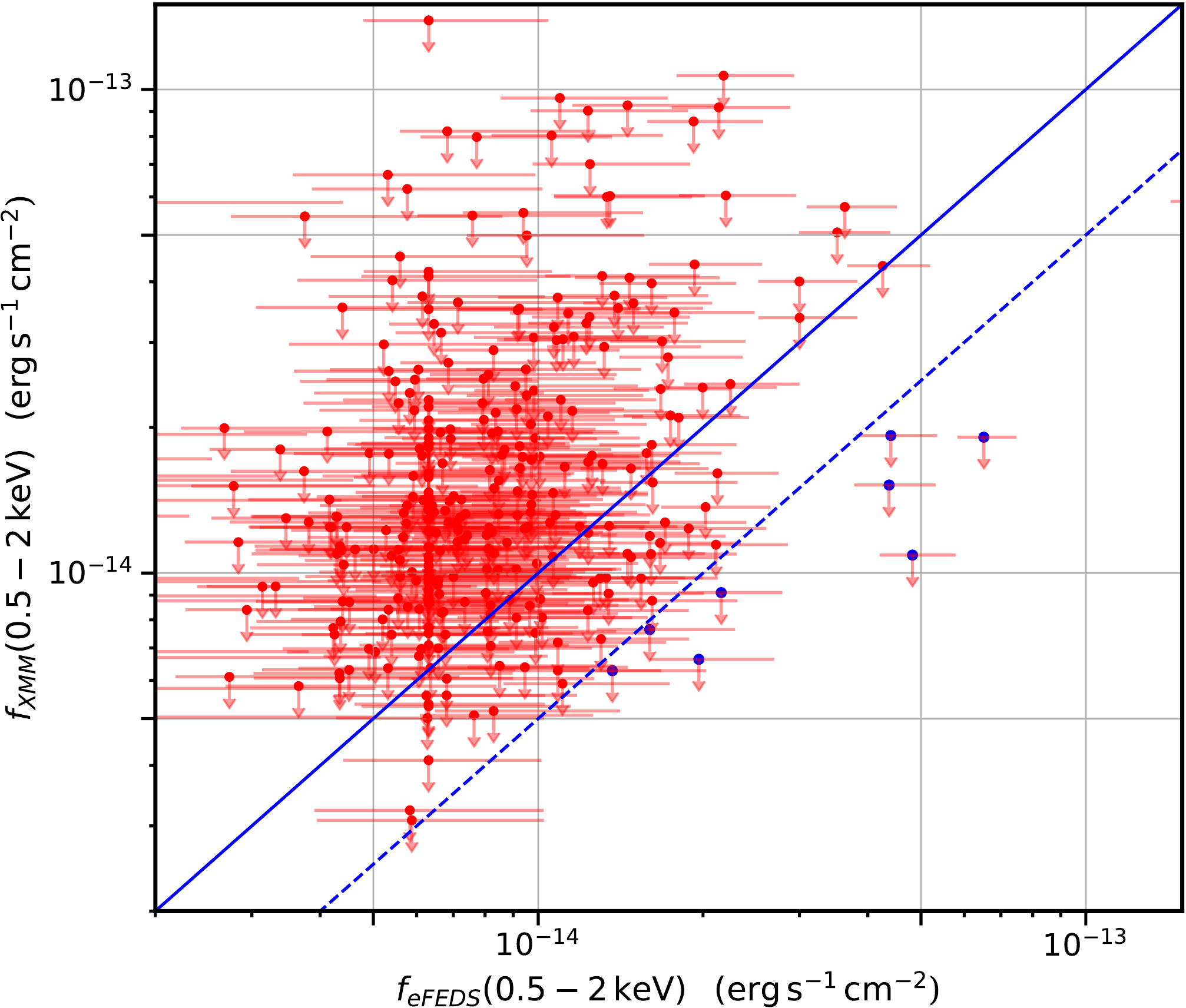}
   \caption{eFEDS 0.5-2\,keV flux vs XMM-ATLAS flux upper-limit ($3\sigma$ confidence) in the 0.5-2\,keV. Each data point on this plot corresponds to an eFEDS source detection without counterpart in the eFEDS source catalogue. For these sources the $3\sigma$ flux upper limits are estimated from the XMM-ATLAS observations and plotted on the vertical axis. The blue solid line indicates the one--to--one relation. The blue dotted line corresponds to an offset by a factor of two in flux relative to the one-to-one relation, i.e. sources with XMM-ATLAS $3\sigma$ flux upper limits that are two times fainter than the XMM-ATLAS source flux. Sources below that line are highlighted by blue colour. Their properties are listed in Table \ref{tab:age:ATLASUPPER}.}
    \label{fig:age:ATLASUPPER}
\end{figure}

\begin{figure}
\includegraphics[width=\columnwidth]{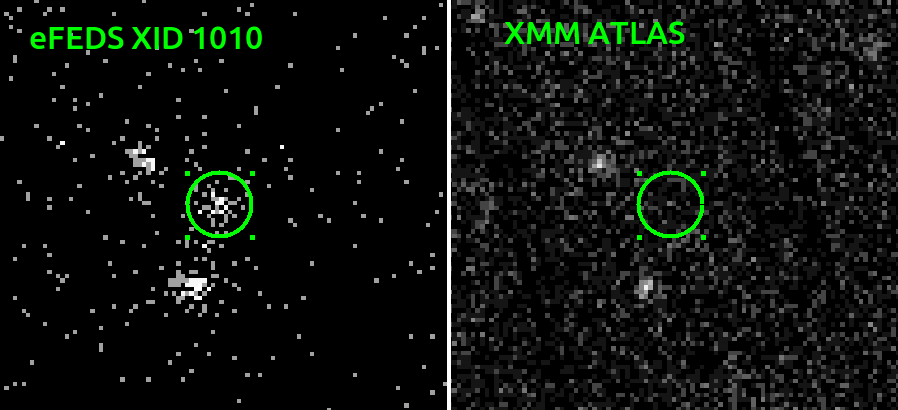}
    \caption{eROSITA and XMM post-stamp images of an eFEDs X-ray source (see Table \ref{tab:age:ATLASUPPER}), which has no X-ray counterpart in the XMM-ATLAS survey. The green circles mark the position of the source in the two observations and have 30\,arcsec radius.}
    \label{fig:age:ATLASUPPER_EXAMPLE}
\end{figure}

Next, eFEDS X-ray sources that are not detected in the XMM-ATLAS observations are considered, defined as eROSITA sources that have no XMM-ATLAS counterpart within 30\,arcsec. For this class of sources a $3\sigma$ upper limit is estimated for their XMM-Newton fluxes in the 0.5-2 keV band. The upper limit calculations is based on aperture photometry, as explained above. The radius of the extraction cell corresponds to the 80\% EEF of the XMM-Newton PSF. The comparison between the eROSITA fluxes and XMM upper limits is shown in Fig.~\ref{fig:age:ATLASUPPER}. Table~\ref{tab:age:ATLASUPPER} presents the sample of sources with flux ratio between the eFEDS and XMM epochs $\ga 2$. Fig.~\ref{fig:age:ATLASUPPER_EXAMPLE} 
shows the XMM-ATLAS and eROSITA/eFEDS X-ray post-stamp images of one of the sources presented in Table~\ref{tab:age:ATLASUPPER}.

\begin{table*}
\begin{tabular}{c cc cc cc l l}
\hline
XID & RA$_X$ & DEC$_X$ & $f_{eFEDS}$ & $f_{XMM}$ & RA$_{opt}$ & DEC$_{opt}$ & $z$ & NW\\
(1) & (2) & (3) & (4) & (5) & (6)  & (7) & (8) & (9) \\
       \hline
188 & 09h06m10.08s &-00d45m53.5s & $1.8^{+0.2}_{-0.2}\times10^{-13}$ & $2.2^{+1.1}_{-0.5}\times10^{-14}$ & 09h06m10.03s &  -00d45m53.5s & 0.223    &  SE\\    
290 & 09h06m12.01s & +01d44m12.0s & $1.4^{+0.2}_{-0.1} \times 10^{-13}$ & $2.3^{+0.4}_{-0.3}\times 10^{-14}$ & 09h06m12.03s & +01d44m11.9s & 0.588 &  SE\\ 
365 & 09h07m12.43s & +00d35m32.4s &  $7.5^{1.2}_{0.8}\times10^{-14}$ & $9^{+3}_{-2}\times10^{-15}$ & 09h07m12.40s & +00d35m32.8s & --               &  LE\\ 
1707 & 09h00m32.3s & +00d37m55.14s & $4.9^{+0.9}_{-0.6}\times 10^{-14}$ & $8.0^{+3}_{2}\times10^{-15}$ & 09h00m32.04s & +00d37m55.2s & 0.7811      &  SE\\ 
4037 & 09h05m01.07s & -00d35m51.5s & $2.1^{+0.7}_{-0.4}\times10^{-14}$ & $1.4^{+0.1}_{-0.1}\times10^{13}$ & 09h05m01.13s &-00d35m50.1s &            &  SG \\ 
5649 & 09h01m59.09s &+00d23m15.3s & $1.0^{+0.5}_{-0.2}\times10^{-14}$   & $7.7^{+0.6}_{-0.5}\times10^{-14}$ & 09h01m58.89s & +00d23m13.9s & 0.196  &  SE\\ 
7260 & 09h02m16.79s & +00d02m49.8s & $8^{+5}_{-2} \times 10^{-15}$ & $5.3^{0.7}_{-0.5}\times10^{-14}$ & 09h02m16.64s & +00d02m48.3s & --            &  LE\\ 
\hline
     \end{tabular}
    \caption{XMM-ATLAS and eFEDS source associations with flux ratio between the two surveys $>6$. The columns are (1): eFEDS X-ray source identification number; (2, 3): eFEDS X-ray sky coordinates, right ascension and declination in J2000; (4) eFEDS flux in the 0.5-2 keV spectral band in units of $\rm erg\, s^{-1}\,cm^{-2}$; (5): the XMM-ATLAS flux in the 0.5-2.0\,keV energy interval in units of $\rm erg\, s^{-1}\,cm^{-2}$; (6, 7): Legacy survey DR8 optical coordinates of the eFEDS source counterpart in J2000 XMM flux, (8): spectroscopic redshift if available. One of the sources is a spectroscopically confirmed Galactic star. 
    (9) NW = NWAY; classification based on \cite{Salvato2021}
    }
    \label{tab:age-ero-atlas-det}
\end{table*}

\begin{table*}
\begin{tabular}{c cc cc c c }
\hline
XID & RA$_X$ & DEC$_X$ & $f_{XMM}$ & $f_{eFEDS}$ & $z$ \\
 -  &  -     &   -    &  ($\rm erg\,s^{-1}\,cm^{-2}$) &  ($\rm erg\,s^{-1}\,cm^{-2}$)  \\
(1) & (2) & (3) & (4) & (5) & (6) \\
       \hline
23776019 & 09h01m13.68s & +01d33m52.2s & $2.7^{+0.4}_{-0.3}\times10^{-14}$ & $<1.04\times10^{-14}$ & 0.569  (phot)   \\ 
23777100 & 09h03m05.87s & +01d48m52.6s & $2.5^{+0.5}_{-0.3}\times10^{-14}$ & $<1.18\times10^{-14}$ &  0.444 (spec)   \\ 
24289026 & 09h05m51.61s & +00d22m07.4s & $3.6^{+0.5}_{-0.4}\times10^{-14}$ & $<1.57\times10^{-14}$ & 0.226 (phot)    \\ 
24544014 & 09h03m14.25s & -00d16m52.8s & $5.2^{+0.6}_{-0.5}\times10^{-14}$ & $<1.05\times10^{-14}$ & 0.341 (phot)    \\ 
24544082 & 09h01m29.50s & +00d14m11.1s & $3.5^{+0.5}_{-0.4}\times10^{-14}$ & $<7.62\times10^{-15}$ &  -              \\
24545009 & 09h08m49.75s & -00d18m44.7s & $2.5^{+0.4}_{-0.3}\times10^{-14}$ & $<1.03\times10^{-14}$ & 0.219 (spec)    \\ 
\hline
     \end{tabular}
    \caption{XMM-ATLAS source detections without counterparts in the eFEDS and flux ratio between the XMM flux and the eFEDS $3\sigma$ upper limit $>2$. The columns are (1): XMM-ATLAS X-ray source identification number; (2, 3): XMM ATLAS X-ray sky coordinates, right ascension and declination, in J2000; (4) XMM flux in the 0.5-2 keV spectral band; (5): eFEDS $3\sigma$ flux upper-limit in the 0.5-2.0\,keV energy interval; (6): SDSS photometric or spectroscopic redshift within 6\,arcsec from the XMM position..
    }
    \label{tab:age:EFEDSUPPER}
\end{table*}

\begin{table*}
\begin{tabular}{cccc cccl l}
\hline
XID & RA$_X$ & DEC$_X$ & $f_{eFEDS}$ & $f_{XMM}$ & RA$_{opt}$ & DEC$_{opt}$ & $z$ & NW \\
(1) & (2) & (3) & (4) & (5) & (6) & (7) & (8) & (9) \\
       \hline
258 & 09h07m52.20s & +01d29m43.6s & $1.6^{+0.2}_{-0.1}\times10^{-13}$ & $<5.9\times10^{-14}$ ($^\star$)& 09h07m52.22s & +01d29m44.4s & 0.102  & SE \\
822 & 09h03m38.75s & +00d53m54.1s & $6.5^{+0.9}_{-0.6}\times10^{-14}$ & $<1.9\times10^{-14}$ & 09h03m38.79s & +00d53m54.2s &     --           & LE \\
1010 & 09h04m58.35s &-00d40m49.9s & $4.3^{+0.9}_{-0.5}\times10^{-14}$ & $<1.5\times10^{-14}$ & 09h04m58.51s &-00d40m47.7s & --                & LE \\
1676 & 09h03m57.51s & -00d07m38.1s & $4.8^{+0.9}_{-0.6}\times10^{-14}$ & $<1.1\times10^{-14}$ & 09h03m57.3s & -00d07m38.5s & --               & LE \\
1842 & 09h08m24.76s & +01d19m19.5s & $4.4^{+0.9}_{-0.6}\times10^{-14}$ & $<1.9\times10^{-14}$ ($^{\star}$) & 09h08m24.66s & +01d19m19.3s & -- & LE \\
4167 & 09h03m10.99s & +00d41m54.3s & $2.1^{+0.6}_{-0.3}\times10^{-14}$ & $<9.1\times10^{-15}$ & 09h03m11.14s & +00d41m54.2s & 0.301           & SE \\ 
5862 & 09h06m51.14s & -00d02m05.5s & $1.9^{+0.7}_{-0.3}\times10^{-14}$ & $<6.6\times10^{-15}$ & 09h06m51.15s &-00d02m03.5s & --               & LE \\
8350 & 09h09m01.65s & -00d07m38.9s & $1.5^{+0.6}_{-0.3}\times10^{-14}$ & $<7.6\times10^{-15}$ & 09h09m01.639s & -00d07m38.7s & --             & LE \\
10271 & 09h07m02.22s & +00d55m52.3s & $1.3^{+0.6}_{-0.2}\times10^{-14}$ & $<6.2\times10^{-15}$ & 09h07m02.17s & +00d55m47.5s & --             & LE \\
\hline
\multicolumn{8}{l}{$^1$eFEDS position is close to the edge of the XMM field of view.}
     \end{tabular}
    \caption{eFEDS source detections without counterparts in the XMM-ATLAS survey and flux ratio between the eFEDS flux and the XMM-ATLAS $3\sigma$ upper limit $>2$. The columns are (1): eFEDS X-ray source identification number; (2, 3): eFEDS X-ray sky coordinates, right ascension and declination in J2000; (4) eFEDS flux in the 0.5-2 keV spectral band in units of $\rm erg\, s^{-1}\,cm^{-2}$; (5): XMM-ATLAS $3\sigma$ flux upper-limit in the 0.5-2.0\,keV energy interval in units of $\rm erg\, s^{-1}\,cm^{-2}$. Positions close to the edge of the XMM-Newton field of view are marked with ($^\star$); (6, 7): Legacy survey DR8 optical coordinates of the eFEDS source counterpart in J2000 XMM flux, (8): spectroscopic redshift if available, (9) NW = NWAY; classification based on \cite{Salvato2021}.
    }
    \label{tab:age:ATLASUPPER}
\end{table*}

\section{Summary}

We have studied time variability of the eFEDS X-ray sources both
on short time scales within the actual eFEDS observations as well
as on longer time scales by comparison to the ROSAT all-sky and
XMM-ATLAS surveys.  About half of the variable objects have been detected for the first time at X-rays with eROSITA.
The number of sources for which significant soft band 
variability within the eFDS observation
has been found is relatively low (65 sources), the majority of
these being late-type stars of spectral type
K and M (c.f. Tab.~\ref{tab:ids}).  The eFEDS survey scanning strategy is very sensitive to stellar flaring events with decay time of a few thousands seconds where up to 6 consecutive data points with time bin sizes of 100 seconds are covered, followed a time gap of about 3600 seconds, up to about 22000 seconds. This is more efficient compared to the eROSITA survey and the ROSAT all-sky survey observations. Our
fits to the eFEDS light curves for the most highly variable sources reveal some extreme stellar flare properties. 
We have cross-matched the 2RXS catalogue with the eFEDS source catalogue and find variability up to a factor of 10 for most of the sources. 12 sources show variability factors above six.
The number of XMM-Atlas detections and eFEDS 3$\sigma$ upper limits is six. Nine sources have eFEDS source detections without counterparts in the XMM-Atlas survey. 
Based on the variability studies in the eFEDS field we expect about 20000 variable sources after completion of the eROSITA survey outside the ecliptic pole regions.

The different time samplings of eFEDS and the all-sky surveys will
qualitatively influence their related variability information.
The scan rate in the all-sky survey is about 7 times faster than
in the eFEDS field scan (90 arcsec/s compared to 13.15 arcsec/s),
and thus typical FOV passing times will be 7 times shorter
(40\,s for central passage).
The survey rate (progression in ecliptic longitude from one
survey great circle to the next) is about 14\,arcmin/4\,h for the eFEDS region
(giving about 4 - 5 passages over a source during one all-sky survey).
However, the strength of the eROSITA all-sky survey observations will
be the comparison between 8 different epochs and the capability to
analyse long-term variability over a 4 years period. 

\begin{acknowledgements}

We thank the anonymous referee for their careful reading of the submitted manuscript, and for their very helpful comments and suggestions.
M.K. acknowledges support by DFG grant KR 3338/4-1.
This work is based on data from eROSITA, the soft X-ray instrument aboard SRG, a joint Russian-German science mission supported by the Russian Space Agency (Roskosmos), in the interests of the Russian Academy of Sciences represented by its Space Research Institute (IKI), and the Deutsches Zentrum für Luft- und Raumfahrt (DLR). The SRG spacecraft was built by Lavochkin Association (NPOL) and its subcontractors, and is operated by NPOL with support from the Max Planck Institute for Extraterrestrial Physics (MPE).
The development and construction of the eROSITA X-ray instrument was led by MPE, with contributions from the Dr. Karl Remeis Observatory Bamberg \& ECAP (FAU Erlangen-Nuernberg), the University of Hamburg Observatory, the Leibniz Institute for Astrophysics Potsdam (AIP), and the Institute for Astronomy and Astrophysics of the University of Tübingen, with the support of DLR and the Max Planck Society. The Argelander Institute for Astronomy of the University of Bonn and the Ludwig Maximilians Universität Munich also participated in the science preparation for eROSITA."
The eROSITA data shown here were processed using the eSASS/NRTA software system developed by the German eROSITA consortium.

\end{acknowledgements}

\bibliographystyle{aa}
\bibliography{paper_main.bbl}

\end{document}